\documentclass[11pt]{article}
\usepackage[utf8]{inputenc}
\usepackage{geometry}
\usepackage{authblk}
\usepackage{setspace}
\usepackage{booktabs}
\usepackage{amssymb}
\usepackage{amsmath}
\usepackage{amsthm}
\usepackage{setspace}
\allowdisplaybreaks[4]
\usepackage{slashed}
\usepackage{multirow}
\usepackage{newunicodechar}
\usepackage{graphicx,color,epsfig}
\usepackage{cite}
\usepackage[colorlinks=true,linkcolor=red,citecolor=blue]{hyperref}
\usepackage{placeins}

\begin{document}
\title{Correlated PQCD Analysis of the Semileptonic Decays $\overline{B}^0 \to D^{(*)+}\ell^-\bar{\nu}_\ell$ and the Nonleptonic Decays $ \overline{B}^0 \to D^{(*)+}\pi^-$}
\author{Mao-Jing Liu}
\author{Ying Li\footnote{liying@ytu.edu.cn}}
\author{Zhi-Tian Zou}
\affil{\it Department of Physics, Yantai University, Yantai 264005,China}
\maketitle
\vspace{0.2cm}
\begin{abstract}
We present a unified analysis of $\overline{B}^0 \to D^{(*)+}\ell^-\bar{\nu}_\ell$ and $\overline{B}^0 \to D^{(*)+}\pi^-$ decays using the perturbative QCD (PQCD) approach. The $B \to D^{(*)}$ transition form factors are calculated at low $q^2$ and extrapolated to the high-$q^2$ region using the latest lattice QCD results via a model-independent $z$-expansion. This hybrid method provides a precise form factor description across the full kinematic range. We then predict the branching fractions and the lepton flavor universality ratios $R(D) = 0.336^{+0.014}_{-0.013}$ and $R(D^*) = 0.271^{+0.010}_{-0.010}$, which are consistent with the latest experimental averages. Furthermore, we perform a correlated study of the nonleptonic $ \overline{B}^0 \to D^{(*)+}\pi^-$ decays, calculating both factorizable and nonfactorizable amplitudes. To reduce hadronic uncertainties, we introduce and calculate the differential ratio $R^{(*)}_{\pi/\ell}(q^2)$, defined between nonleptonic and semileptonic decay rates, providing a sensitive test of factorization and possible new physics effects. The predictions presented here can be directly tested in ongoing Belle II and LHCb experiments.
\end{abstract}
\section{Introduction}
Semileptonic $b \to c \ell^- \bar{\nu}_\ell$ ($\ell = e, \mu, \tau$) transitions provide a clean environment to study weak interactions in the presence of strong dynamics. In particular, the exclusive decays $\overline{B}^0 \to D^{(*)+} \ell^- \bar{\nu}_\ell$ are essential for determining the Cabibbo-Kobayashi-Maskawa (CKM) matrix element $|V_{cb}|$ \cite{Cabibbo:1963yz, Kobayashi:1973fv, Bernlochner:2022ywh,Gambino:2019sif}, for testing the structure of charged-current interactions, and for validating nonperturbative QCD calculations of heavy-to-heavy form factors. These channels also play a crucial role in testing the Standard Model (SM) and probing possible new physics (NP) effects. Of particular interest are the lepton flavor universality (LFU) ratios, defined as
\begin{eqnarray}\label{RDdef}
R(D^{(*)}) = \frac{\mathcal{B}(B \to D^{(*)} \tau \bar{\nu}_\tau)}{\mathcal{B}(B \to D^{(*)} \mu \bar{\nu}_\mu)}\,,
\end{eqnarray}
which are predicted with high precision in the SM owing to significant cancellation of hadronic uncertainties. These observables are sensitive to charged-current interactions mediated by $W$ bosons, and any deviation from SM expectations could point to new interactions, such as scalar or vector currents from physics beyond the SM \cite{Crivellin:2018yvo, Duan:2024ayo,Iguro:2024hyk,Capdevila:2023yhq}.

On the theory side, the SM predictions for $R(D)$ and $R(D^*)$ have reached percent-level precision. The Heavy Flavour Averaging Group (HFLAV) \cite{HeavyFlavorAveragingGroupHFLAV:2024ctg} quotes (spring 2025)
\begin{align}
	R(D)_{\text{SM}} = 0.296 \pm 0.004, \,\,\,\,\
	R(D^*)_{\text{SM}} = 0.254 \pm 0.005, \label{HFLAV}
\end{align}
which are obtained from global fits that combine lattice QCD inputs \cite{FlavourLatticeAveragingGroupFLAG:2024oxs, Martinelli:2023fwm, Ray:2023xjn} at high-$q^2$ with experimental form factor determinations at low-$q^2$. These results rely on extrapolating the form factors to the low-$q^2$ region, making them sensitive to the choice of parameterization, with uncertainties dominated by form-factor precision at nonzero recoil.

Experimentally, $R(D)$ and $R(D^*)$ have been measured by the BaBar, Belle, and LHCb collaborations using different techniques, including hadronic and semileptonic tagging. The latest HFLAV world averages (spring 2025) \cite{HeavyFlavorAveragingGroupHFLAV:2024ctg} are:
\begin{align}
	R(D)_{\text{exp}} = 0.347 \pm 0.025, \,\, \,\,
	R(D^*)_{\text{exp}} = 0.288 \pm 0.012. \label{HFLAV}
\end{align}
These results include both statistical and systematic uncertainties. Combining $R(D)$ and $R(D^*)$ measurements and taking into account their correlation, the overall discrepancy with the SM is approximately $3.8\sigma$. While this is smaller than the earlier $\sim 4\sigma$ tension, it remains one of the most persistent hints of LFU violation in the $B$ sector. Forthcoming high-luminosity data from Belle~II and the upgraded LHCb are expected to substantially reduce the uncertainties on $R(D)$ and $R(D^*)$. With $50\mathrm{ab}^{-1}$ of data, Belle~II aims for a precision of about $2\%$ on both observables~\cite{Belle-II:2018jsg}, while LHCb Run 3 will further improve these measurements by increasing the instantaneous luminosity fivefold and enhancing trigger efficiency across most modes by a factor of two~\cite{Kutsenko:2025ahl}.


While semileptonic decays $\overline{B}^0 \to D^{(*)+} \ell^- \bar{\nu}_\ell$ are free from hadronic final-state interactions in the leptonic sector, their amplitudes depend directly on the hadronic transition form factors $\overline{B}^0 \to D^{(*)+}$. From a theoretical perspective, the results quoted in Eq.~(\ref{HFLAV}) are not strictly theory predictions, as they partially rely on experimental input. A more rigorous strategy is to determine the low-$q^2$ form factors from first-principles or QCD-based approaches such as perturbative QCD (PQCD) or light-cone sum rules (LCSR) and then extrapolate them to the high-$q^2$ region, where lattice QCD and heavy-quark effective theory (HQET) provide precise results \cite{Bernlochner:2022ywh, Boyd:1995sq, Caprini:1997mu}. This leads to a fully theory-driven description of the form factors across the entire kinematic domain. In recent years, the low-$q^2$ form factors of $\overline{B}^0 \to D^{(*)+}$ have been computed within LCSR \cite{Faller:2008tr, Wang:2017jow, Gao:2021sav, Cui:2023jiw}, and when combined with lattice inputs, the results agree well with Eq.~(\ref{HFLAV}). Nevertheless, it remains valuable to study them within the PQCD framework, especially at low-$q^2$. This region corresponds to the large-recoil limit, where the hard-scattering mechanism dominates and PQCD offers a systematically improvable framework with Sudakov resummation suppressing endpoint singularities. Moreover, PQCD predictions at low-$q^2$ provide an independent cross-check of nonperturbative methods, thereby helping to quantify model dependence and systematic uncertainties. Finally, the low-$q^2$ form factors serve as crucial boundary conditions for parameterizations over the full kinematic range. This strategy was initially explored in Refs.~\cite{Fan:2015kna, Hu:2019bdf}. Therefore, we shall calculate the form factors in the low-$q^{2}$ region using the PQCD framework and extrapolate them to the full kinematic range by incorporating the latest lattice QCD results \cite{Na:2015kha,Harrison:2023dzh, Aoki:2023qpa}. This combined strategy yields a form-factor description that is both theoretically consistent and firmly rooted in first-principles calculations, providing reliable coverage across the entire phase space. Such a framework enhances the precision of phenomenological studies, including global determinations of CKM matrix elements and stringent tests of LFU.

In addition, the nonleptonic decay modes $\overline{B}^0 \to D^{(*)+} \pi^-$ share the same heavy-to-heavy transition currents as their semileptonic counterparts, making them an excellent testing ground for exploring the correlation between these two classes of decays. However, their amplitudes also receive contributions from nonfactorizable diagrams, which cannot be reliably calculated using naive factorization but can be systematically evaluated within the PQCD framework. Notably, PQCD allows for a consistent treatment of both factorizable and nonfactorizable contributions, providing a unique opportunity to link the hadronic dynamics probed in semileptonic decays with those governing nonleptonic processes. In the PQCD approach to calculating these decays, the nonperturbative wave functions of heavy mesons play a crucial role and represent the primary source of theoretical uncertainty. We aim to establish correlations between these decays, which helps to reduce hadronic uncertainties. Any significant and consistent deviation between such correlated theoretical predictions and experimental data could serve as a signal of NP beyond the SM. In this work, we also perform a correlated PQCD analysis of $\overline{B}^0 \to D^{(*)+} \ell^- \bar{\nu}_\ell$ and $\overline{B}^0 \to D^{(*)+} \pi^-$ decays by combining lattice QCD results at high-$q^2$ with PQCD calculations at low-$q^2$. Our goal is to provide a unified description of these processes and to assess their implications for LFU tests and hadronic dynamics.

This paper is organized as follows. In Sec.~\ref{sec:pqcd}, we outline the theoretical framework and summarize the PQCD formalism employed for both semileptonic and nonleptonic $B$ decays, including the treatment of factorizable and nonfactorizable contributions. In Sec.~\ref{sec:semileptonic}, we present our calculation of the $\overline{B}^0 \to D^{(*)+}$ form factors in the low-$q^2$ region within PQCD and discuss their matching to the latest lattice QCD results at high $q^2$. Then, we apply these form factors to evaluate the branching ratios, differential distributions, and polarization observables for $\overline{B}^0 \to D^{(*)+} \ell^- \bar{\nu}_\ell$, with particular emphasis on LFU-sensitive quantities. In Sec.~\ref{sec:nonleptonic}, we analyze the nonleptonic decays $\overline{B}^0 \to D^{(*)+} \pi^-$ in PQCD, highlighting the uncertainties of $B$ meson wave function. In Sec.~\ref{sec:correlation}, we shall explore the correlation between the semileptonic results and nonleptonic decays. Finally, Sec.~\ref{sec:summary} contains our conclusions and outlook.

\section{Framework} \label{sec:pqcd}
\subsection{Brief Review of PQCD Approach}
The PQCD approach based on the $k_T$ factorization framework has been developed and extensively applied to nonleptonic $B$ meson decays \cite{Keum:2000wi,Lu:2000em,Ali:2007ff}. In this approach, the decay amplitude is factorized into contributions from soft ($\Phi$), hard ($H$), and harder ($C$) dynamics, each characterized by distinct energy scales. We take $\overline B \to D^+ \pi^-$ as an example, and write its amplitude as the convolution \cite{Li:1992nu}
\begin{equation}
	\mbox{Amplitude}
	\sim \int\!\! d^4k_1 d^4k_2 d^4k_3\
	\mathrm{Tr} \bigl[ C(t) \Phi_B(k_1) \Phi_{D}(k_2) \Phi_\pi(k_3)
	H(k_1,k_2,k_3, t) \bigr],
	\label{eq:convolution1}
\end{equation}
where $k_i$ denote the momenta of the light quarks within each meson, and $\mathrm{Tr}$ stands for the trace over Dirac and color indices. The Wilson coefficient $C(t)$ arises from short-distance radiative corrections and incorporates the harder dynamics at scales above the $B$ meson mass $m_B$. It governs the renormalization group evolution of local four-Fermi operators from the electroweak scale $m_W$ down to the intermediate scale $t \sim \mathcal{O}(\sqrt{\bar{\Lambda} m_B})$, where $\bar{\Lambda} \equiv m_B - m_b$. The hard kernel $H$ describes the interaction between the four-quark operator and the spectator quark mediated by a hard gluon with virtuality $q^2 \sim \bar{\Lambda} m_B$, capturing the perturbative dynamics at the scale $\mathcal{O}(\sqrt{\bar{\Lambda} m_B})$. Hence, $H$ can be calculated perturbatively. The wave function $\Phi_M$ characterizes the hadronization process of a quark-antiquark pair into the meson $M$. While $H$ depends on the specific decay process, $\Phi_M$ is universal and process-independent. By determining $\Phi_M$ from other decay channels, we can make quantitative predictions for the processes considered here.

For simplicity, we consider the $B$ meson at rest. It is convenient to adopt light-cone coordinates $(p^+, p^-, \mathbf{p}_T)$ to describe the meson momenta, where
\begin{eqnarray}
p^\pm = \frac{1}{\sqrt{2}} (p^0 \pm p^3),\,\,\,{\bf p}_T = (p^1, p^2).
\end{eqnarray}
In this notation, the momenta of the $B$, $D$, and $\pi$ mesons are taken as
\begin{eqnarray}
P_1 = \frac{m_B}{\sqrt{2}} (1,1,{\bf 0}_T),\,\,\,
P_2 = \frac{m_B}{\sqrt{2}} (1,r^2,{\bf 0}_T),\,\,\,
P_3 = \frac{m_B}{\sqrt{2}} (0,1-r^2,{\bf 0}_T),
\end{eqnarray}
where $r = m_D / m_B$, and the pion mass $m_\pi$ is neglected. Let $k_1$, $k_2$, and $k_3$ denote the momenta of the light (anti)quarks inside the $B$, $D$, and $\pi$ mesons, respectively. We choose
\begin{eqnarray}
k_1 = (x_1 P_1^+,0,{\bf k}_{1T}),\,\,\,
k_2 = (x_2 P_2^+,0,{\bf k}_{2T}),\,\,\,
k_3 = (0, x_3 P_3^-,{\bf k}_{3T}),
\end{eqnarray}
where $x_i$ are the corresponding longitudinal momentum fractions and $\mathbf{k}_{iT}$ are the transverse momenta. Integrating over $k_1^-$, $k_2^-$, and $k_3^+$ in Eq.~(\ref{eq:convolution1}) yields
\begin{multline}
	\text{Amplitude} \sim \int
	dx_1 dx_2 dx_3
	b_1 db_1 b_2 db_2 b_3 db_3 \\
	\times \mathrm{Tr} \bigl[ C(t)  \Phi_B(x_1, b_1)  \Phi_{D}(x_2, b_2)
	\Phi_\pi(x_3, b_3)  H(x_i, b_i, t) S_t(x_i)  e^{-S(t)} \bigr],
	\label{eq:convolution2}
\end{multline}
where ${\bf b}_i$ is the coordinate conjugate to ${\bf k}_{iT}$, and $t$  denotes the largest energy scale in the hard kernel $H$, expressed as a function of $x_i$ and $b_i$. The large logarithms $\ln(m_W/t)$, arising from QCD radiative corrections to the four-quark operators, are resummed into the Wilson coefficients $C(t)$. Large double logarithms $\ln^2 x_i$ in the longitudinal direction are resummed via threshold resummation~\cite{Li:2001ay}, resulting in the function $S_t(x_i)$, which smooths the end-point singularities in $x_i$. The Sudakov factor $e^{-S(t)}$ contains two types of logarithms: a single logarithm $\ln(tb)$ from the renormalization of ultraviolet divergences, and a double logarithm $\ln^2 b$ from the overlap of collinear and soft gluon corrections. This Sudakov suppression effectively damps soft contributions~\cite{Li:1997un}, ensuring that the perturbative calculation of $H$ remains valid at the intermediate scale $\mathcal{O}(m_B)$.

\subsection{Wave Function of Meson}
The meson wave functions $\Phi_{M,\alpha\beta}$, serving as the primary nonperturbative inputs, can be decomposed in terms of their spin structures. With Dirac indices $\alpha$ and $\beta$, they are expanded in the complete basis of 16 independent Dirac matrices: $1_{\alpha\beta}$, $\gamma^\mu_{\alpha\beta}$, $\sigma^{\mu\nu}_{\alpha\beta}$, $(\gamma^\mu\gamma_5)_{\alpha\beta}$, and $\gamma_{5\alpha\beta}$. For a heavy pseudoscalar meson such as $B$ or $D$, only the $(\gamma^\mu\gamma_5)_{\alpha\beta}$ and $\gamma_{5\alpha\beta}$ components contribute at leading order, yielding
\begin{equation}
	\Phi_{M,\alpha\beta} = \frac{i}{\sqrt{2N_c}}
	\left\{
	(\not \! P_M \gamma_5)_{\alpha\beta} \phi_M^A
	+ \gamma_{5\alpha\beta} \phi_M^P
	\right\},
\end{equation}
where $N_c = 3$ is the number of colors, $P_M$ is the meson momentum, and $\phi_M^{A,P}$ are Lorentz-scalar distribution amplitudes. In HQET, $\phi_B^P \simeq m_B \phi_B^A$, so the $B$-meson wave function takes the form
\begin{equation}
	\Phi_{B,\alpha\beta}(x,b) = \frac{i}{\sqrt{2N_c}}
	\left[	(\not \! P_1 \gamma_5)_{\alpha\beta}+ m_B \gamma_{5\alpha\beta}
	\right]  \phi_B(x,b). \label{waveB}
\end{equation}
Given that $\phi_B(x,b)$ is sharply peaked in the small-$x$ region, we adopt the parametrization
\begin{equation}
\phi_B(x,b) = N_B  x^2 (1-x)^2
\exp \left[
-\frac{m_B^2 x^2}{2 \omega_b^2}
- \frac{1}{2} (\omega_b b)^2
\right],
\end{equation}
as in Refs.~\cite{Keum:2000wi, Lu:2000em}, which provides an excellent fit to $B\to K\pi$ and $B\to\pi\pi$ data. It is noted that in recent years the high power contribution of $B$-meson light-cone distribution amplitude have been studied extensively, for examples in Refs.~\cite{Wang:2019msf, Wang:2016qii, Wang:2018wfj, Wang:2024wwa}.

For a fast-moving pseudoscalar meson such as the $D$ meson, the wave function is described by three Lorentz-scalar distribution amplitudes $\phi$, $\phi_p$, and $\phi_\sigma$ \cite{Ball:1998je}:
\begin{gather}
    \langle D^+(P)|{\bar c}(0)\gamma_\mu \gamma_5 d(z)| 0 \rangle \simeq
    - i f_{D} P_\mu \int_0^1\!\! dx\; e^{i x P\cdot z}\,\phi(x),
    \label{pv} \\
    \langle D^+(P)|{\bar c}(0)\gamma_5 d(z)|0 \rangle =
    - i f_{D} m_{0D} \int_0^1\!\! dx\; e^{i x P\cdot z}\,\phi_p(x),
    \label{ps} \\
    \langle D^+(P)|{\bar c}(0)\gamma_5 \sigma_{\mu\nu} d(z)|0 \rangle =
    \frac{i}{6} f_{D} m_{0D} \left(1-\frac{m_{D}^2}{m_{0D}^2} \right)
    (P_\mu z_\nu-P_\nu z_\mu)
    \int_0^1\!\! dx\; e^{i x P\cdot z}\,\phi_\sigma(x),
    \label{pt}
\end{gather}where $m_{0D} = m_{D}^2/(m_c + m_d)$.

In the perturbative calculation, the small difference between the $c$-quark mass and the $D$-meson mass is neglected. Defining $\bar{\Lambda}' \equiv m_{D} - m_c$, we drop terms proportional to $\bar{\Lambda}'/m_{D}$. With this approximation, the contribution from Eq.(\ref{pt}) is suppressed by ${\cal O}(\bar{\Lambda}'/m_{D})$ compared with Eqs.(\ref{pv}) and (\ref{ps}), owing to the factor $1 - m_{D}^2/m_{0D}^2$. Thus, the $\gamma_5 \sigma_{\mu\nu}$ term is omitted in the $D$-meson wave function.

Furthermore, using Eqs.~(\ref{pv}) and (\ref{ps}) together with
\begin{gather}
\frac{\partial}{\partial z_\mu}
\langle D^+(P)|\bar c(0)\gamma_\mu\gamma_5 d(z)|0\rangle
= i m_d \langle D^+(P)|\bar c(0)\gamma_5 d(z)|0\rangle, \\[6pt]
\frac{\partial}{\partial z_\mu}
\langle D^+(P)|\bar c(z)\gamma_\mu\gamma_5 d(0)|0\rangle
= -im_c \langle D^+(P)|\bar c(z)\gamma_5 d(0)|0\rangle.
\end{gather}
and applying the equations of motion, one finds
\begin{equation}
\phi_p(x) = \phi(x) +{\cal O}\left( \frac{\bar{\Lambda}'}{m_{D}} \right).
\end{equation}

Therefore, only a single $D$ meson wave function is involved in our calculations \cite{Keum:2003js,Lu:2002iv},
\begin{eqnarray}
\Phi_{D,\alpha\beta}(x,b)
=\frac{i}{\sqrt{2N_c}}\Big[(\gamma_5\not\!P)_{\alpha\beta} +m_D \gamma_{5\alpha\beta}\Big]
\phi_D(x,b)\;, \label{waveD}
\end{eqnarray}
where the distribution amplitude,
\begin{eqnarray}
\phi_D=\frac{f_D}{2\sqrt{2N_c}}\phi_D^v=\frac{f_D}{2\sqrt{2N_c}}\phi_D^p\;,
\end{eqnarray}
satisfies the normalization,
\begin{eqnarray}
\int_0^1 dx\phi_D(x)=\frac{f_D}{2\sqrt{2N_c}}\;.
\end{eqnarray}
For the purpose of numerical estimate, we adopt the simple model \cite{Keum:2003js,Lu:2002iv,Li:2008tk},
\begin{eqnarray}
\phi_D(x)=\frac{3}{\sqrt{2N_c}}f_D x(1-x)[1+C_D(1-2x)]\;.
\label{phid}
\end{eqnarray}
The free shape parameter $C_D=0.80\pm0.05$ is chosen such that the distribution amplitude $\phi_D$ peaks around $x \sim \bar{\Lambda}/m_D \sim 0.3$. We do not consider the intrinsic $b$ dependence of the $D$ meson wave function, which can be introduced along with more free parameters. Note that Eq.~(\ref{phid}) differs from the one of the Gaussian form proposed in \cite{Li:1999kna, Wu:2025kdc, Zhang:2017rwz}.

Neglecting the $O(\bar\Lambda/m_{D^*})$ contribution, we have the structure for a $D^*$ meson,
\begin{eqnarray}
\Phi_{D^*,\alpha\beta}(x,b)
=\frac{i}{\sqrt{2N_c}}\Big[\not\!\varepsilon_L^*\not \!P+m_{D^*}\Big]_{\alpha\beta}\phi_{D^*}^L(x,b),
\label{dss}
\end{eqnarray}
and the $D^*$ meson distribution amplitude is given as
\begin{eqnarray}
\phi_{D^*}^L(x)=\phi_{D^*}^T(x)=
\frac{3}{\sqrt{2N_c}}f_{D^*}x(1-x)[1+C_{D^*}(1-2x)]\;.
\label{phis}
\end{eqnarray}
Similarly, the free shape parameter $C_{D^*}=0.80\pm0.05$ is expected to take a value, so that  $\phi_{D^*}$ has a maximum at $x\sim \bar\Lambda/m_{D^*}\sim 0.3$.

\subsection{Wave Functions and Decay Constants of Light Pseudoscalar Mesons}
The decay constant of the pseudoscalar meson is defined as:
 \begin{eqnarray}
 \langle \pi^-(P)|\bar d\gamma_{\mu}\gamma_5u|0\rangle=-if_\pi P_{\mu},
 \end{eqnarray}
 with $ f_{\pi} = 131 \mbox{MeV}$. The Lorentz structure of wave function (for out-going state) for a $\pi^-$ meson is
 \begin{align}
\langle \pi^-(P)|\bar{d}_{\alpha}(z)u_{\beta}(0)|0\rangle
=\frac{i}{\sqrt{2N_c}}\int_0^1 dx
e^{ixP\cdot z}\left[\gamma_5\not\!{P}\phi_\pi^A(x)+\gamma_5m_{0\pi}\phi_\pi^P(x)+m_{0\pi}\gamma_5(\not\! v\not\! n-1)\phi_\pi^T(x)\right]_{\beta\alpha}, \label{wavePI}
\end{align}
where $x$ is the momentum fraction carried by the $d$ quark, vector $v$ is parallel to the pion meson momentum $P_3$, and $n$ is just opposite to it. The chiral scale parameter $m_{0\pi}$ is defined as $m_{0\pi}=\frac{m_\pi^2}{m_{u}+m_{d}}$. The distribution amplitudes are expanded by the Gegenbauer polynomials and their expressions are given as \cite{Ball:1998je}:
 \begin{eqnarray}
 &&\phi_\pi^A(x)=\frac{3f_\pi}{\sqrt{2N_c}}x(1-x)\left[1+a_2^AC_2^{3/2}(t)+a_4^AC_4^{3/2}(t)\right],\\
 &&\phi_\pi^P(x)=\frac{f_\pi}{2\sqrt{2N_c}}\left[1+a_2^PC_2^{1/2}(t)+a_4^PC_4^{1/2}(t)\right],\\
 && \phi_\pi^T(x)=-\frac{f_\pi}{2\sqrt{2N_c}}\left[C_1^{1/2}(t)+a_3^TC_3^{1/2}(t)\right],
 \end{eqnarray}
with $t=2x-1$. The coefficients of the Gegenbauer polynomials are \cite{Braun:1988qv,Li:2008ts,Zou:2009zza}
 \begin{eqnarray}
a^A_{2}=0.44\;,\;a^A_{4}=0.25\;,
 a^P_{2}= 0.43\;,\;a^P_{4}=0.09\;,
 a^T_{3}= 0.55\;.
 \end{eqnarray}
\section{The Semi-leptonic Decays $\overline B^0 \to D^{(*)+} \ell^- \bar{\nu}_\ell$} \label{sec:semileptonic}
\subsection{$B \to D^{(*)}$ Form Factors}
The form factors for $\overline B \to D $ transition are defined by \cite{Wang:2012ab}:
\begin{eqnarray}
\left \langle D(P_2) \left | \bar{c}\gamma ^\mu b \right | \overline B(P_1) \right \rangle = f_1(q^2)P_1^\mu + f_2(q^2)P_2^\mu
\end{eqnarray}
Another equivalent definition is
\begin{equation}
\left \langle D(P_2) \left | \bar{c}\gamma ^\mu b \right |  \overline B(P_1) \right \rangle = \left [ (P_1+P_2)^\mu-\frac{m_B^2}{q^2}q^\mu \right ] F_+(q^2)+\frac{m_B^2+m_D^2}{q^2}q^\mu F_0(q^2).
\end{equation}
The relationship between these two sets of form factors can be represented by the following equations
\begin{align}
F_+(q^2) &= \frac{1}{2}[f_1(q^2) + f_2(q^2)], \\
F_0(q^2) &= \frac{1}{2}f_1(q^2)\left[1 + \frac{q^2}{m_B^2 - m_D^2}\right] + \frac{1}{2}f_2(q^2)\left[1 - \frac{q^2}{m_B^2 - m_D^2}\right].
\end{align}

The $B\to D^* $ form factors are defined through the following decompositions of hadronic matrix elements,
\begin{eqnarray}
\langle D^*(P_2,\varepsilon^*)|\bar{c}\gamma^\mu b|\overline B(P_1)\rangle
&=&\frac{2iV(q^2)}{m_B+m_{D^*}}\epsilon^{\mu\nu\rho\sigma}
\varepsilon^*_\nu P_{2} P_{1\sigma}\;,
\label{vf}\\
\langle D^* (P_2,\varepsilon^*)|\bar{c}\gamma^\mu \gamma_5 b
|\overline B(P_1)\rangle
&=&2m_\rho A_0(q^2)\frac{\varepsilon^*\cdot q}{q^2}q^\mu+
(m_B+m_{D^*})A_1(q^2)\left[\varepsilon^{*\mu}-
\frac{\varepsilon^*\cdot q}{q^2}q^\mu\right]
\nonumber\\
& &-A_2(q^2)\frac{\varepsilon^*\cdot q}{m_B+m_{D^*}}
\left[P_1^\mu+P_2^\mu-\frac{m_B^2-m_{D^*}^2}{q^2}q^\mu\right]\;.
\label{af}
\end{eqnarray}

In the $B$-meson rest frame, we denote the momentum of the $B$ menson as $P_1$, and the momentum of the $D^{(*)+}$ meson as $P_2$
 \begin{equation}
P_1 = \frac{m_B}{\sqrt{2}}(1,1,\mathbf{0}_T) \quad P_2 = \frac{r m_B}{\sqrt{2}}(\eta^+,\eta^-,\mathbf{0}_T),
\end{equation}
with the factors $\eta^\pm=\eta\pm\sqrt{\eta^2-1}$ is defined in terms of the parameter $\eta=\frac{1}{2r}[1+r^2-\frac{q^2}{m_B^2}]$. Here, the ratio $r=m_{D^{(*)}} /m_B$, and $q=P_1-P_2$ is the lepton-pair momentum. The longitudinal polarize vector $\varepsilon_L$ and the transverse polarization vector $\varepsilon_T$ of the $D^*$ meson are given by $\varepsilon_L=\frac{1}{\sqrt 2}(\eta^+,-\eta^-,\mathbf{0}_T )$, $\varepsilon_T=(0,0,\mathbf{1}_T)$. The momenta of the spectator quarks inside the $B$ and $D^{(*)}$ mesons are chosen as
\begin{equation}
k_1=(0,x_1P_1^-,\mathbf{k}_{1T} )\quad k_2=(x_2P_2^+, x_2P_2^-,\mathbf{k}_{2T}).
\end{equation}

\begin{figure}[phtb]
		\centering
		\includegraphics[width=1.0\textwidth]{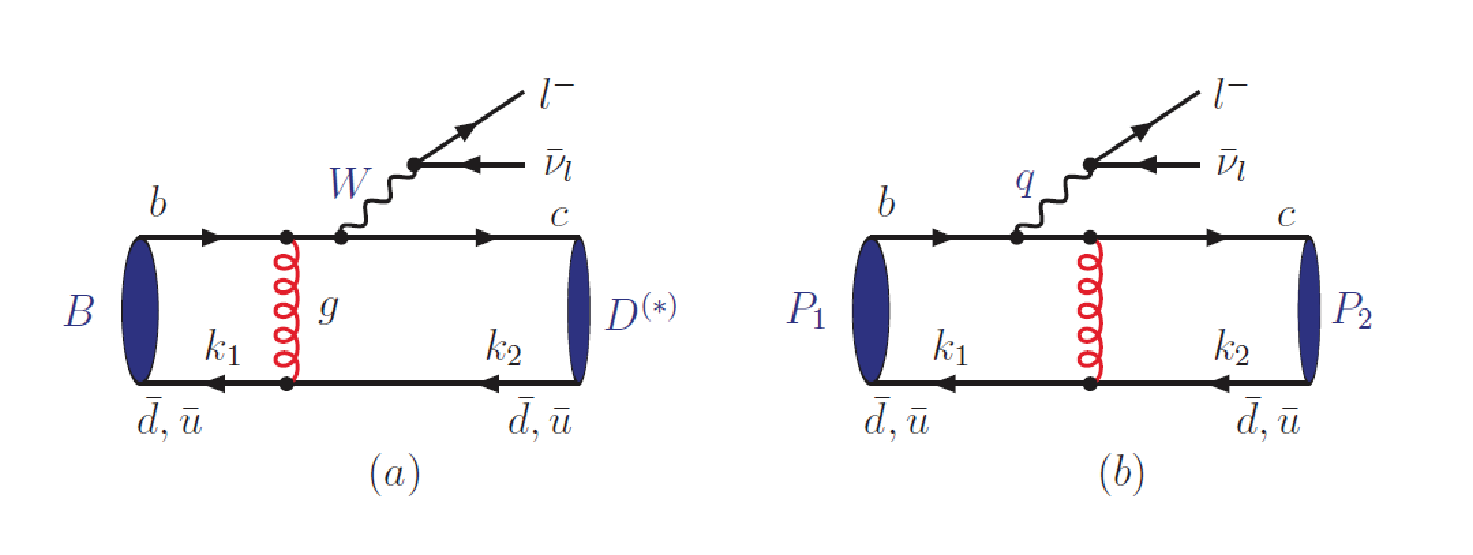}
		\caption{The lowest order Feynman diagrams for the semileptonic decays $\overline{B}^0 \to D^{(*)+}\ell^-\bar{\nu}_\ell$ in PQCD.}
\label{fig:fig1}
\end{figure}

In the PQCD framework, the leading-order Feynman diagrams for the semileptonic decays  $\overline{B}^0 \to D^{(*)+} \ell^- \bar{\nu}_{\ell}$ are shown in Fig.~\ref{fig:fig1}.   For the $\overline B \to D$ transition form factors, we evaluate the two relevant diagrams and derive the expressions for $f_1$ and $f_2$ as
\begin{align}
f_1(q^2) &= 8\pi m_B^2 C_F \int_{0}^{1} dx_1 dx_2 \int_{0}^{1/\Lambda} b_1 db_1 b_2 db_2 \, \phi_B(x_1, b_1)\,  \phi_D(x_2, b_2)\nonumber \\
& \times \Big\{\left( 2r(1 - rx_2)\right)  \, h_1(x_1, x_2, b_1, b_2) \,  \alpha_s(t_1) \, \exp[-S_B(t_1) -S_{D}(t_1)]  \nonumber\\
& + \left( 2r(2r_c-r) + x_1 r \left( -2 + 2\eta + \sqrt{\eta^2 - 1} - \frac{2\eta}{\sqrt{\eta^2 - 1}} + \frac{\eta^2}{\sqrt{\eta^2 - 1}} \right) \right)  \nonumber\\
& \times h_2(x_1, x_2, b_1, b_2) \, \alpha_s(t_2)  \, \exp[-S_B(t_2) -S_{D}(t_2)] \Big\},
\label{f1}\\
f_2(q^2) &= 8\pi m_B^2 C_F \int_{0}^{1} dx_1 dx_2 \int_{0}^{1/\Lambda} b_1 db_1 b_2 db_2\,  \phi_B(x_1, b_1)\,  \phi_D(x_2, b_2) \nonumber\\
&\times \Big\{ \left(2-4x_2r(1-\eta)\right)  \, h_1(x_1, x_2, b_1, b_2) \,\alpha_s(t_1)\,  \exp[-S_B(t_1) -S_{D}(t_1)]  \nonumber\\
& + \left( 4r-2r_c-x_1+\frac{x_1}{\sqrt{\eta^2 - 1}}\left( 2-\eta\right) \right)\, h_2(x_1, x_2, b_1, b_2) \, \alpha_s(t_2) \, \exp[-S_B(t_2) -S_{D}(t_2)]  \Big\},
\label{f2}
\end{align}
with $r_c=m_c/m_{B}$ and color factor $ C_F=4/3$. The hard scales $t_i$  are chosen as the largest scale of the virtuality of the internal particles
\begin{eqnarray}
t_1=\max\{m_{B}\sqrt{x_2 r \eta^+}, 1/b_1, 1/b_2\},\quad\quad
t_2=\max\{m_{B}\sqrt{x_1 r \eta^+},1/b_1, 1/b_2\}.
\end{eqnarray}
The hard functions $h_{1,2}(x_i,b_i)$ come from the Fourier transform and can be written as \cite{Wang:2013ix} :
\begin{align}
h_1(x_1, x_2, b_1, b_2) &= K_0(\beta_1 b_1) \Big\{ \theta(b_1 - b_2) I_0(\alpha_1 b_2) K_0(\alpha_1 b_1)  + \theta(b_2 - b_1) I_0(\alpha_1 b_1) K_0(\alpha_1 b_2) \Big\} \, S_t(x_2),\\
h_2(x_1, x_2, b_1, b_2) &= K_0(\beta_2 b_1) \Big\{ \theta(b_1 - b_2) I_0(\alpha_2 b_2) K_0(\alpha_2 b_1) + \theta(b_2 - b_1) I_0(\alpha_2 b_1) K_0(\alpha_2 b_2) \Big\} \, S_t(x_1),
\end{align}
where $K_0$ and $I_0$ are the modified Bessel functions of the second and first kind, respectively. The parameters $\alpha_i$ and $\beta_i$ are given by
\begin{equation}
\alpha_1=m_B\sqrt{x_2r\eta^+} \quad \alpha_2=m_B\sqrt{x_1r\eta^+} \quad
\beta_1=\beta_2=m_B\sqrt{x_1x_2r\eta^+}
\end{equation}
Here, $S_t(x_i)$ is the threshold resummation factor, and the Sudakov factor $S_{B,D}(t)$ are also referred to Ref.~\cite{Wang:2013ix}.

Similarly, for the $B \to D^*$ transition form factors, we compute the corresponding leading-order diagrams and obtain the expressions for $V$ and $A_{0,1,2}$ as
\begin{align}
V(q^2) &= 8\pi m_B^2 C_F \int_{0}^{1} dx_1 dx_2 \int_{0}^{1/\Lambda} b_1 db_1 b_2 db_2\,  \phi_B(x_1, b_1)\,  \phi^T_{D^*}(x_2, b_2) \, (1+r) \nonumber\\
& \times \Big\{\left(1 - rx_2\right)  \, h_1(x_1, x_2, b_1, b_2) \, \alpha_s(t_1) \,  \exp[-S_B(t_1) -S_{D^*}(t_1)]  \nonumber\\
& + \left( r+ \frac{x_1}{2 \sqrt{\eta^2-1}} \right)\, h_2(x_1, x_2, b_1, b_2) \,  \alpha_s (t_2)) \,  \exp[-S_B(t_2) -S_{D^*}(t_2)] \Big\},
\label{V}\\
A_0(q^2) &= 8\pi m_B^2 C_F \int_{0}^{1} dx_1 dx_2 \int_{0}^{1/\Lambda} b_1 db_1 b_2 db_2 \, \phi_B(x_1, b_1) \, \phi^L_{D^*}(x_2, b_2) \nonumber\\
&  \times \Big\{ \left(1+r-rx_2(2+r-2 \eta)\right)  \,h_1(x_1, x_2, b_1, b_2) \,\alpha_s(t_1) \exp[-S_B(t_1) -S_{D^*}(t_1)] \nonumber\\
&  + \left( r^2+r_c+ \frac{x_1}{2 }+ \frac{\eta x_1}{2 \sqrt{\eta^2 -1}} + \frac{rx_1}{2 \sqrt{\eta^2 -1}} \big(1-2 \eta (\eta+\sqrt{\eta^2 -1}) \big) \right)\nonumber\\
&  \times h_2(x_1, x_2, b_1, b_2) \, \alpha_s(t_2) \, \exp[-S_B(t_2) -S_{D^*}(t_2)]  \Big\},
\label{a0}\\
A_1(q^2) &= 8\pi m_B^2 C_F \int_{0}^{1} dx_1 dx_2 \int_{0}^{1/\Lambda} b_1 db_1 b_2 db_2 \phi_B(x_1, b_1) \, \phi^T_{D^*}(x_2, b_2) \, \frac{r}{1+r}\nonumber\\
& \times \Big\{2\left(1 + \eta -2rx_2 + r \eta x_2\right)  \, h_1(x_1, x_2, b_1, b_2) \, \alpha_s(t_1) \, \exp[-S_B(t_1) -S_{D^*}(t_1)]  \nonumber\\
& + \left(2r_c + 2 \eta r - x_1 \right)\, h_2(x_1, x_2, b_1, b_2) \, \alpha_s(t_2) \, \exp[-S_B(t_2) -S_{D^*}(t_2)] \Big\},
\label{a1}\\
A_2(q^2) &= \frac{(1+r)^2(\eta -r)}{2r(\eta^2-1)} \, A_1(q^2) - 8\pi m_B^2 C_F \int_{0}^{1} dx_1 dx_2 \int_{0}^{1/\Lambda} b_1 db_1 b_2 db_2 \, \phi_B(x_1, b_1)\,  \phi^L_{D^*}(x_2, b_2) \nonumber\\
& \times \frac{1+r}{\eta^2-1} \, \Big\{ \left[(1+ \eta )(1-r)-rx_2\left(1-2r+ \eta (2+r-2 \eta )\right)\right]\nonumber\\
& \times h_1(x_1, x_2, b_1, b_2) \, \alpha_s(t_1) \exp[-S_B(t_1) -S_{D^*}(t_1)]  \nonumber\\
& + \left(r+r_c(\eta-r)- \eta r^2 - \frac{x_1}{2}(\eta +r)+x_1(\eta r- \frac{1}{2})\sqrt{\eta^2-1}] \right)\nonumber\\
& \times h_2(x_1, x_2, b_1, b_2) \, \alpha_s(t_2) \, \exp[-S_B(t_2) -S_{D^*}(t_2)]  \Big\}.\label{a2}
\end{align}
It is shown that the form factor $A_0(q^2)$ is determined exclusively by the longitudinal wave function, while $V(q^2)$ and $A_1(q^2)$ are governed solely by the transverse wave function. In contrast, $A_2(q^2)$ receives contributions from both longitudinal and transverse components.

\begin{figure}[ptbp]
        \includegraphics[width=0.5\linewidth]{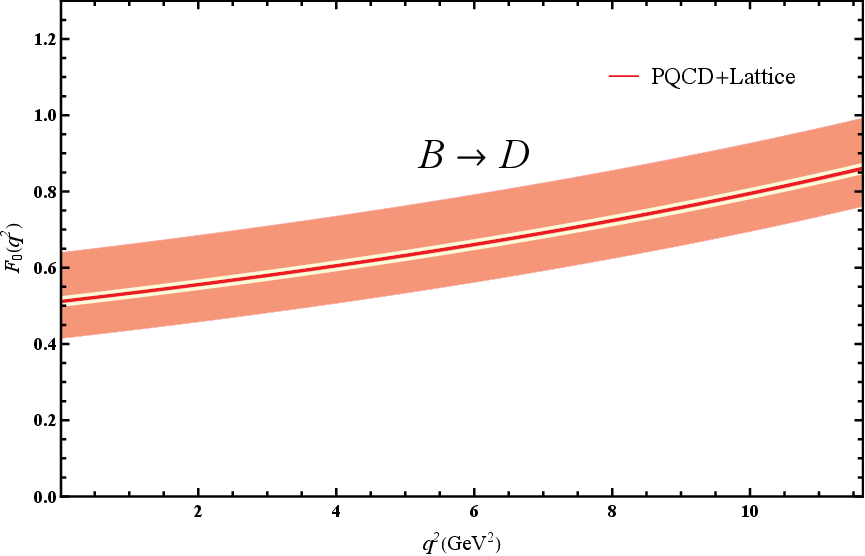}
        \includegraphics[width=0.5\linewidth]{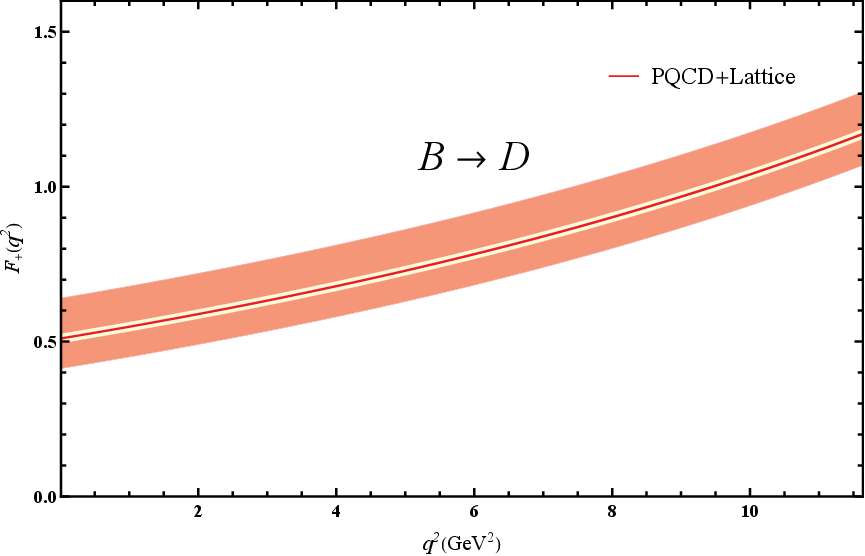}
        \includegraphics[width=0.5\linewidth]{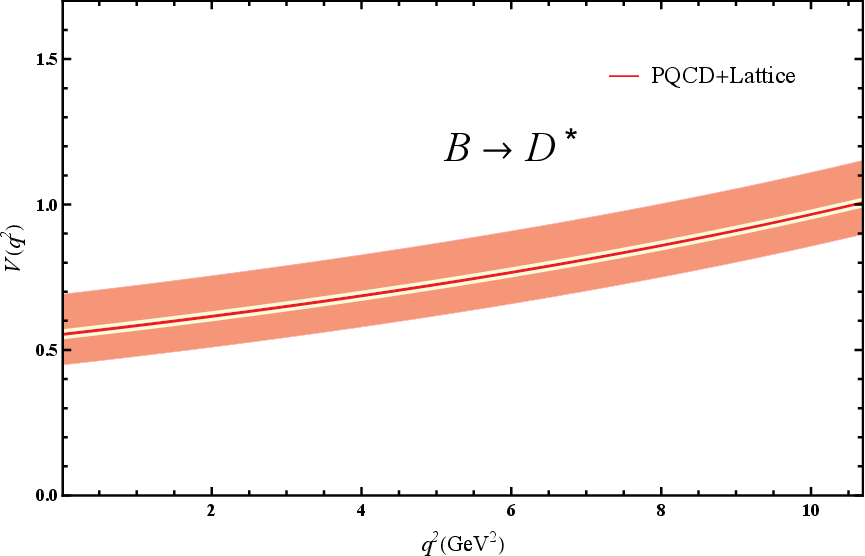}
        \includegraphics[width=0.5\linewidth]{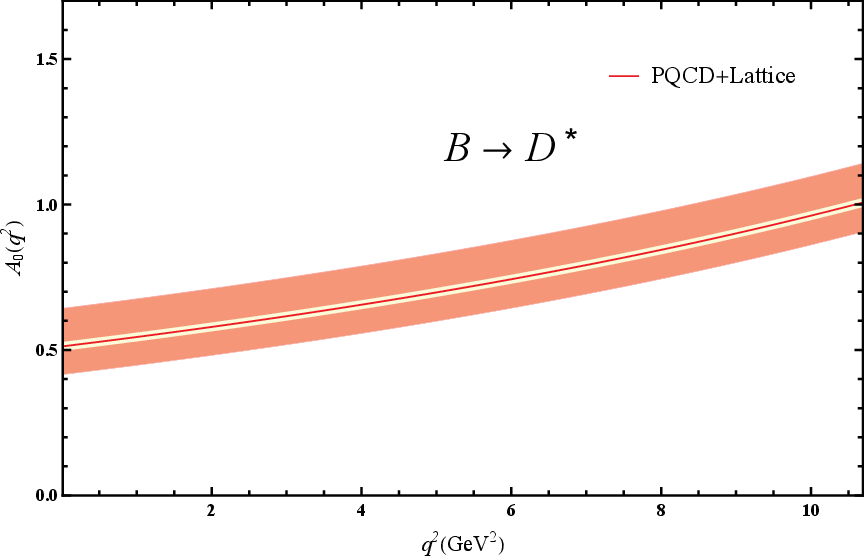}
        \includegraphics[width=0.5\linewidth]{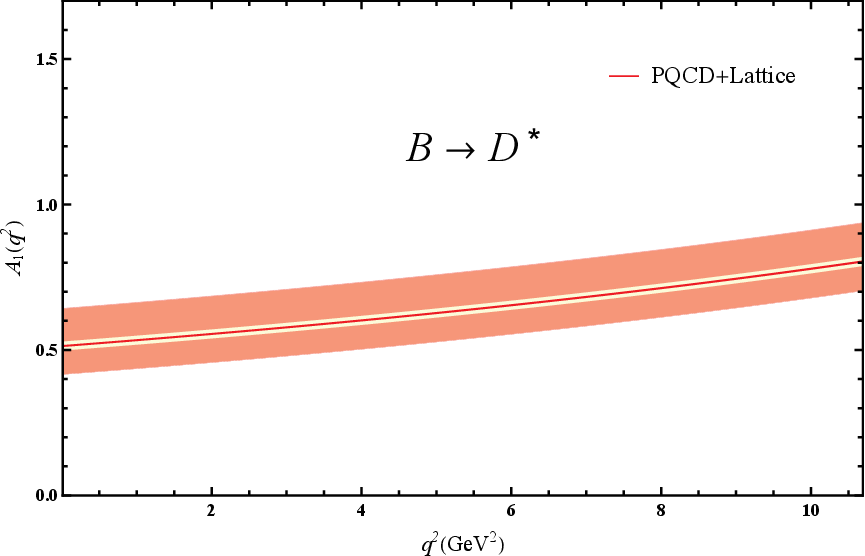}
        \includegraphics[width=0.5\linewidth]{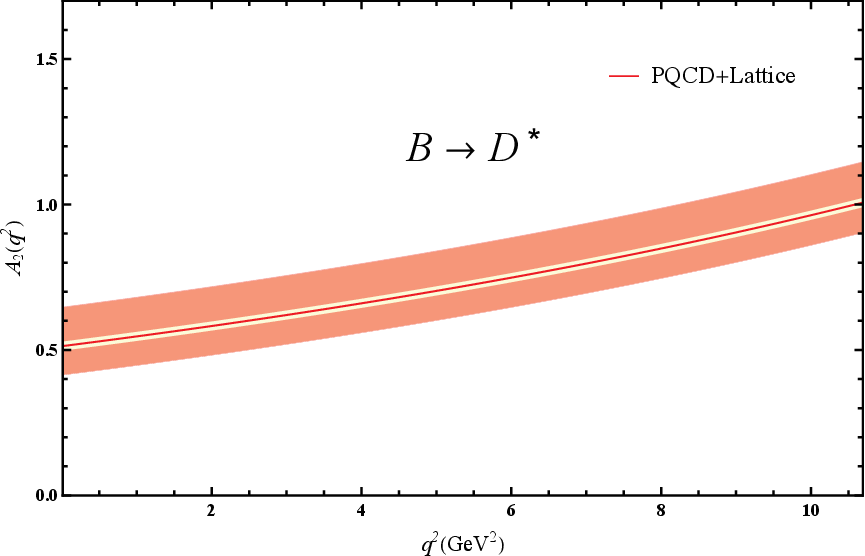}
    \caption{The theoretical predictions for the $q^2$-dependence of the form factors for $B \to (D,D^*)$ transitions in the PQCD approach (the blue solid curves) , the ``PQCD + Lattice" method (the red solid curves), error varying with $\omega_B$ (shaded in lightred) and with $cd$ (shaded in lightyellow)}. \label{fig:formfactor}
\end{figure}

As discussed above, it is widely accepted that lattice QCD provides reliable predictions for the relevant form factors in the high-$q^2$ region, while PQCD calculations are applicable at low $q^2$. Accordingly, we first evaluate the form factors for $B \to D^{(*)}$  transition at ten points in the low- $q^2$ region,$0\le q^2 \le m_{\mu} ^2$ , using the PQCD approach. The numerical results at $q^2=0$ are summarized in Table.~\ref{fitresults}. By combining these PQCD results with the lattice QCD inputs \cite{FlavourLatticeAveragingGroupFLAG:2024oxs,Na:2015kha,Harrison:2023dzh}, also listed in Table~\ref{fitresults}, we perform an extrapolation of the form factors from the low- $q^2$ to the high-$q^2$ region, thereby obtaining their behavior across the entire kinematic range. The extrapolation indeed exhibits parametric dependence. In current work, rather than adopting the pole model commonly used in the literature \cite{Wang:2008xt, Fan:2015kna}, we employ the Bourrely-Caprini-Lellouch (BCL) parametrization  \cite{Boyd:1995cf, Bourrely:2008za, Hu:2019qcn},
\begin{equation}
    f_{i}(q^2)=\frac{1}{1-q^2 / m_{R}^{2}} \sum_{k=0}^{1} \alpha_{k}^{i} z^{k}\left(q^2, t_{0}\right)=\frac{1}{1-q^2 / m_{R}^{2}}\left(\alpha_{0}^{i}+\alpha_{1}^{i} \frac{\sqrt{t_{+}-q^2}-\sqrt{t_{+}-t_{0}}}{\sqrt{t_{+}-q^2}+\sqrt{t_{+}-t_{0}}}\right). 
\label{eq:bcl}
\end{equation}
In above, we only retain the first two terms of the series in the parameter $z$, with $t_0 = (m_B + m_{D^{(*)}})(\sqrt{m_B} - \sqrt{m_{D^{(*)}}})^2$ and $ t_{+}=(m_B + m_{D^{(*)}})^2$.  All relevant resonance masses $m_R$ \cite{Leljak:2019eyw} are  given in Table.~\ref{fitresults}, together with the fitted parameters $\alpha_0$, $\alpha_1$ from Eq. (\ref{eq:bcl}). The predicted form factors in a full $q^2$ range are shown in Fig.~\ref{fig:formfactor}.

\begin{table}[ptbp]
\centering
    \caption{Summary of the BCL fit for $B \to D^{(*)}$ form factors.}
    \label{fitresults}
    \begin{tabular}{cccccc}
     \hline
     \hline
    Form factor & $m_{R}({\rm GeV})$ & $\alpha_0$ & $\alpha_1$ & $q^2=0$ & $q^2=q_{max}^2$ \\
     \hline
      $F_{+}$ & 6.34 & 0.67 & -4.98& $0.51$ & $1.17$\\
      $F_{0}$ & 6.71 & 0.58 & -1.96& $0.51$ & $0.86$\\
      $V_{0}$ & 6.34 & 0.66 & -3.18& $0.55$ & $1.01$\\
      $A_{0}$ & 6.28 & 0.63 & -3.78& $0.51$ & $1.01$\\
      $A_{1}$ & 6.75 & 0.57 & -1.74& $0.51$ & $0.80$\\
      $A_{2}$ & 6.75 & 0.66 & -4.39& $0.51$ & $1.01$\\
      \hline
      \end{tabular}
\end{table}

\subsection{The Semileptonic Decays $\overline{B}^0 \to D^{(*)+}\ell^-\bar{\nu}_\ell$}
For the semileptonic decays $\overline{B}^0 \to D^{(*)+}\ell^-\bar{\nu}_{\ell}$ , the quark level transitions are $b \to c \ell^- \bar{\nu}_\ell$ decays with the effective Hamiltonian
\begin{eqnarray}
\mathcal{H}_{eff}(b \to c \ell^- \bar{\nu}_\ell)=\frac{G_F}{\sqrt{2}}V_{cb}\bar{c}\gamma_{\mu}(1-\gamma_5)b\cdot\bar{\ell}\gamma^\mu(1-\gamma_5)\nu_\ell,
\end{eqnarray}
where $G_F$ is the Fermi-coupling constant, $V_{cb}=0.0416$ is the CKM matrix element.

The differential decay widths of the semileptonic decays  $\overline{B}^0 \to D^+\ell^-\bar{\nu}_{\ell}$ can be written as \cite{Li:2008tk}
\begin{align}
\frac{d \Gamma (\overline{B}^0 \to D^{+}\ell^-\bar{\nu}_{\ell} )}{dq^2} &=\frac{G_F^2 \left |V_{cb} \right| ^2}{192 \pi^3 m_B^3}\left(1-\frac{m_\ell^2}{q^2}\right)^2 \frac{\lambda^{1/2}(m_B^2,m_{D}^2,q^2)}{2q^2}\nonumber\\
& \times \left [3m_\ell^2(m_B^2-m_D^2)^2\left | F_0(q^2) \right |^2 + (m_\ell^2+2q^2)\lambda(m_B^2,m_{D}^2,q^2)\left | F_+(q^2) \right |^2 \right], \label{DBDlnu}
\end{align}
where $\lambda(x,y,z)=x^2+y^2+z^2-2(xy+yz+zx)$ is the triangular function. For  $\overline{B}^0 \to D^{*+}\ell^-\bar{\nu}_{\ell}$ decays, the differential decay widths can be written as
\begin{align}
\frac{d \Gamma_L (\overline{B}^0 \to D^{*+}\ell^-\bar{\nu}_{\ell})}{dq^2}  &= \frac{G_F^2 \left |V_{cb} \right |^2}{192 \pi^3 m_B^3}\left(1-\frac{m_\ell^2}{q^2}\right)^2 \frac{\lambda^{1/2}(m_B^2,m_{D^*}^2,q^2)}{2q^2} \Bigg\{3m_\ell^2 \lambda(m_B^2,m_{D^*}^2,q^2)A^2_0(q^2)\nonumber\\
&
 + \frac{m_\ell^2+2q^2}{4m_{D^*}^2 } \left[ (m_B^2-m_{D^*}^2-q^2)(m_B+m_{D^*})A_1(q^2)
 -\frac{\lambda(m_B^2,m_{D^*}^2,q^2)}{m_B+m_{D^*}}A_2(q^2) \right]^2 \Bigg\}
\label{dgl}\\
\frac{d \Gamma_\pm (\overline{B}^0 \to D^{*+}\ell^-\bar{\nu}_{\ell})}{dq^2}  &= \frac{G_F^2 \left |V_{cb} \right |^2}{192 \pi^3 m_B^3}\left(1-\frac{m_\ell^2}{q^2}\right)^2 \frac{\lambda^{3/2}(m_B^2,m_{D^*}^2,q^2)}{2}\nonumber\\
& \times  \left(m_\ell^2+2q^2\right) \left[\frac{V(q^2)}{m_B+m_{D^*}} \mp \frac{(m_B+m_{D^*})A_1(q^2)}{\sqrt{\lambda(m_B^2,m_{D^*}^2,q^2)}} \right]^2
\label{dgp}
\end{align}
The total differential decay widths is then given as
\begin{equation}
\frac{d \Gamma }{dq^2} =\frac{d \Gamma_L}{dq^2}+\frac{d \Gamma_+}{dq^2}+\frac{d \Gamma_-}{dq^2}. \label{DBDslnu}
\end{equation}

\begin{figure}[phtb]
    \centering
         \includegraphics[width=0.48\linewidth]{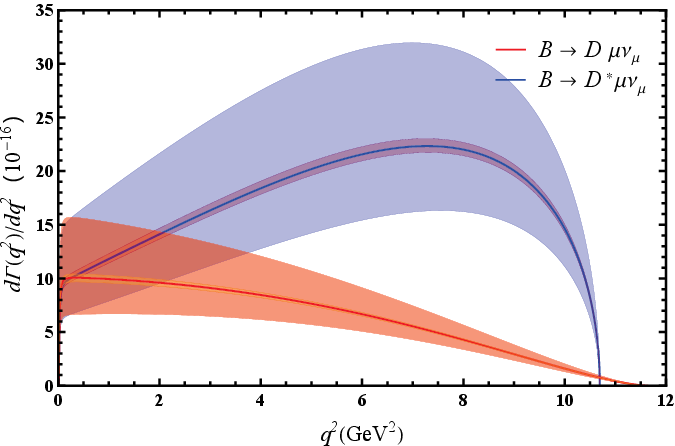}
         \includegraphics[width=0.48\linewidth]{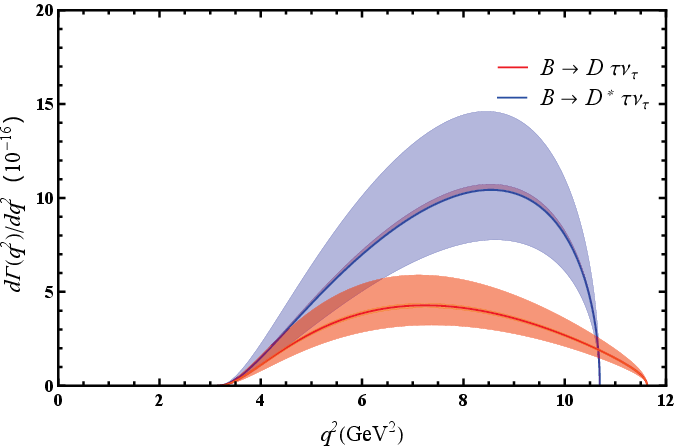}
	\caption{PQCD predictions for the $q^2$-dependence of the differential decay widths $d\Gamma(B \to D^{(*)} \ell \bar{\nu}_\ell)/dq^2$. The  solid curves denote the central values, while the larger uncertainties stem from the parameter $\omega_B$, and the thinner uncertainties from the parameter $C_{D^{(*)}}.$}\label{fig:K_Kt}
\end{figure}

Using the Eqs.~(\ref{DBDlnu}) and (\ref{DBDslnu}), we present the PQCD predictions for the semileptonic decays $\overline{B}^0 \to D^{(*)+}\ell^- \bar{\nu}_\ell$ with $\ell=\mu,\tau$. The $q^2$-dependence of the differential decay widths is displayed in  Fig.~\ref{fig:K_Kt}. The left panel shows the prediction for the muon channel $\overline B^0 \to D^{(*)+} \mu^- \bar{\nu}_\mu$, while the right panel (b) corresponds to the tau channel $\overline B^0 \to D^{(*)+} \tau^- \bar{\nu}_\tau$. The solid curves represent the central predictions of the PQCD approach, and the accompanying uncertainty bands are highly informative: the larger uncertainties stem from the parameter $\omega_B$, which is related to the shape of the $B$-meson distribution amplitude and is a dominant source of theoretical error, while the thinner uncertainties from the parameter $C_{D^{(*)}}$ indicate a comparatively better-controlled systematic. The different slopes  of the two distributions highlight the role of spin and polarization effects in $\overline B^0 \to D^{(*)}$ transitions, reflecting the distinct underlying form-factor dynamics.

After integrating over $q^2$, we then obtain the total branching fractions:
\begin{align}
  {\cal B}(\overline{B}^0 \to D^{+}\mu^-\bar{\nu}_{\mu})  &= \left(1.65^{+0.71+0.05}_{-0.45-0.05}\right)\times 10^{-2}, \\
  {\cal B}(\overline{B}^0 \to D^{*+}\mu^-\bar{\nu}_{\mu}) &= \left(4.33^{+1.99+0.15}_{-1.24-0.12}\right)\times 10^{-2}, \\
  {\cal B}(\overline{B}^0 \to D^{+}\tau^-\bar{\nu}_{\tau})  &= \left(5.54^{+2.08+0.14}_{-1.35-0.15}\right)\times 10^{-3}, \\
  {\cal B}(\overline{B}^0 \to D^{*+}\tau^-\bar{\nu}_{\tau}) &= \left(11.75^{+4.88+0.35}_{-3.09-0.31}\right)\times 10^{-3}.
\end{align}
The quoted uncertainties reflect the combined effects of $\omega_B$ and $C_{D^{(*)}}$, with the variation of $\omega_B$ again providing the dominant contribution. The relative uncertainty reaches $\mathcal{O}(20\%)$ in some channels, underscoring the importance of precise determinations of meson distribution amplitudes. In the experimental side, the branching fractions of these decays are measured to be as \cite{ParticleDataGroup:2024cfk}:
\begin{align}
  {\cal B}(\overline{B}^0 \to D^{+}\mu^-\bar{\nu}_{\mu})  &= \left(2.10\pm 0.07\right)\times 10^{-2}, \label{exbr-1} \\
  {\cal B}(\overline{B}^0 \to D^{*+}\mu^-\bar{\nu}_{\mu}) &= \left(4.87\pm0.09\right)\times 10^{-2}, \label{exbr-2}\\
  {\cal B}(\overline{B}^0 \to D^{+}\tau^-\bar{\nu}_{\tau})  &= \left(9.8\pm 2.1\right)\times 10^{-3}, \label{exbr-3}\\
  {\cal B}(\overline{B}^0 \to D^{*+}\tau^-\bar{\nu}_{\tau}) &= \left(1.48\pm0.09\right)\times 10^{-2}.\label{exbr-4}
\end{align}
A comparison between the theoretical predictions and experimental measurements for the $\overline{B}^0 \to D^{(*)} \ell^- \bar{\nu}_{\ell}$ decays reveals a consistent picture. For all four decay channels, the experimental central values lie within the relatively large theoretical uncertainties, indicating a broad consistency between theory and experiment. However, a clear and systematic trend is observed when examining the central values: the experimental measurements are uniformly higher than the theoretical predictions. Specifically, for the $D\mu\nu$, $D^*\mu\nu$, $D\tau\nu$, and $D^*\tau\nu$ modes, the data exceeds the theory by approximately $28\%$, $13\%$, $77\%$, and $26\%$, respectively. Given that an increase in the shape parameter $\omega_B$ suppresses the predicted branching fractions, this consistent upward shift in the data strongly suggests that the experimental measurements favor a smaller value of $\omega_B$ than the reference value of $0.4\rm GeV$ used in the calculation. This provides crucial input for future global fits to precisely determine this non-perturbative QCD parameter.

In our calculation, the LFU ratios are obtained within the SM using PQCD form factors supplemented by lattice inputs, yielding
\begin{equation}
R(D)|_{\rm PQCD } = 0.336^{+0.014}_{-0.013} ,
\qquad
R(D^*) |_{\rm PQCD }= 0.271^{+0.010}_{-0.010}.
\end{equation}
The only significant residual uncertainties in our results arise from the $B$-meson shape parameter $\omega_B$, while the uncertainties from the $D^{(*)}$ meson are negligibly small. Compared with the corresponding branching fractions, the errors in $R(D^{(*)})$ are much smaller due to the partial cancellation of common hadronic inputs, making these ratios particularly robust for phenomenological comparisons. Our predictions are fully consistent with the latest experimental averages and lie somewhat closer to the data than the SM expectations based on lattice results  combined with experimental results shown in Eq.~(\ref{HFLAV}). The comparison between the two SM frameworks highlights the sensitivity of these ratios to the treatment of hadronic form factors, especially at nonzero recoil. Our results therefore provide a complementary SM determination, and further improvements in lattice inputs and PQCD calculations are expected to reduce theoretical uncertainties, enabling a more precise assessment of potential lepton flavor universality violations in future experiments.

From the measured branching fractions given in Eqs.~(\ref{exbr-1})–(\ref{exbr-4}),  we derive the ratios $R(D)=0.467\pm0.101$ and $R(D^*)=0.304\pm0.019$.  These values are noticeably higher than the directly measured HFLAV 2025 averages given in Eq. \eqref{HFLAV}, particularly for $R(D)$, showing a clear difference in the central values while remaining  statistically consistent within the quoted uncertainties.  Compared with the PQCD-based SM predictions obtained in current work, the ratios calculated from the measured branching fractions are again slightly larger but compatible within uncertainties.  Such deviations, though not yet significant, highlight the importance of future high-precision  measurements from Belle-II and LHCb, together with improved theoretical determinations of  hadronic form factors from lattice QCD and PQCD, to further scrutinize possible departures from the SM in semileptonic $B$ decays.

\begin{figure}[phtb]
	\centering
	\includegraphics[width=0.6\textwidth]{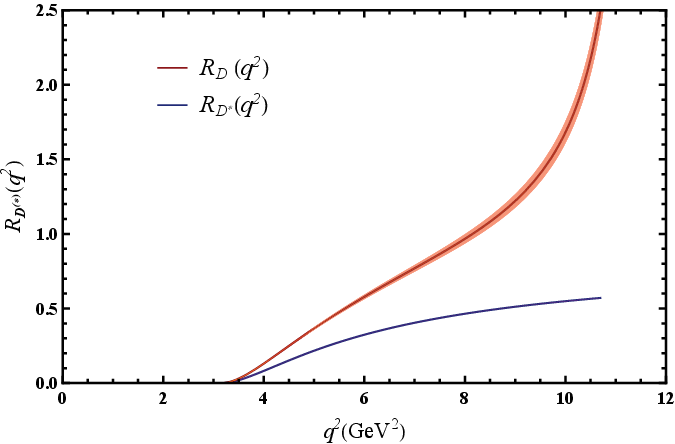}
	\caption{PQCD predictions for the ratios $R_D(q^2)$ and $R_{D^*}(q^2)$ defined in
	Eq.~\eqref{DRDdef}. The bands showing the uncertainties from $\omega_B$.}
	\label{fig:RD}
\end{figure}

In order to further probe LFU, we also define the differential ratios
\begin{eqnarray}\label{DRDdef}
R_{D^{(*)}}(q^2)=
\frac{d\Gamma(\overline B^0 \to D^{(*)+} \tau^- \bar{\nu}_\tau)/dq^2}{d\Gamma(\overline B^0 \to D^{(*)+} \mu^- \bar{\nu}_\mu)/dq^2}\,,
\end{eqnarray}
which provide a $q^2$-dependent measurement of LFU. Unlike the integrated ratios, the differential ratios allow a more detailed comparison between theory and experiment across the entire kinematic range. We plot $R_{D^{(*)}}(q^2)$ to illustrate its $q^2$ dependence in Fig.~\ref{fig:RD}. The residual errors shown in the shaded bands are therefore much smaller than those in the absolute decay widths. We also note that at low $q^2$, the ratios are suppressed due to the reduced phase space available for the $\tau$ channel. As $q^2$ increases, the $\tau$ contribution becomes more prominent, and both ratios rise toward their maxima near the kinematic endpoint. The sharp increase of $R_D(q^2)$ in the high-$q^2$ region reflects the growing impact of the scalar form factor, while $R_{D^*}(q^2)$ shows a more moderate increase, consistent with its dominant vector and axial-vector contributions. Hadronic uncertainties largely cancel in these ratios, making them particularly sensitive probes of potential new physics effects. If future measurements of the differential ratios were to show significant deviations from our predictions, such discrepancies could provide clear evidence for physics beyond the SM.

Compared with previous studies \cite{Fan:2015kna, Hu:2019bdf}, this work introduces several important improvements. First, in the extrapolation of the form factors, we incorporate the latest lattice QCD results \cite{Na:2015kha,Harrison:2023dzh, Aoki:2023qpa}, which provide more precise values near the high-$q^2$ endpoint and thus reduce theoretical uncertainties. Moreover, we employ the model-independent $z$-expansion parametrization instead of the traditional pole model  \cite{Fan:2015kna}, further minimizing the error associated with parametrization dependence. Second, for the $D$-meson wave function, we adopt the one extracted from fits to non-leptonic $B\to DP,DV$ decays \cite{Li:2008ts} rather than assuming a Gaussian-type distribution, leading to a more realistic description of hadronic dynamics. Third, in handling the charm-quark propagator, we avoid the unphysical imaginary parts that appeared in earlier works due to the explicit inclusion of the charm-quark mass in the propagator denominator. In this study, we take the charm-quark mass equal to the $D$-meson mass in the denominator, which naturally removes such unphysical effects. Finally, we use $\Lambda_{\text{QCD}} = 0.25~\text{GeV}$, corresponding to a smaller $\alpha_s$, which improves the convergence of the perturbative series and reduces higher-order corrections, thereby enhancing the stability and reliability of the PQCD calculation.

\section{The Non-leptonic Decays $\overline B^0 \to D^{(*)+} \pi^-$ }\label{sec:nonleptonic}
Since the considered decays $\overline B^0 \to D^{(*)+} \pi^-$ correspond to the $b\to c \bar{u}d$ transition, we describe the effective Hamiltonian as \cite{Buchalla:1995vs}
 \begin{multline}
\mathcal{H}_{eff}(b\to c \bar{u}d)=\frac{G_{F}}{\sqrt{2}}V_{ud}^{*}V_{cb}
\Big[C_{1}(\mu )\bar{d}_{\alpha} \gamma_{\mu} (1-\gamma_{5}) u_{\beta} \gamma^{\mu} (1-\gamma_{5}) \bar{c}_{\beta} b_{\alpha}  \\
+C_{2}(\mu) \bar{d}_{\alpha} \gamma_{\mu} (1-\gamma_{5}) u_{\alpha} \bar{c}_{\beta}\gamma^{\mu} (1-\gamma_{5})  b_{\beta} \Big], \label{eff}
\end{multline}
where $\alpha$ and $\beta$ are the color indices, $V_{ud}^{*}V_{cb}$ is the product of the CKM matrix elements , and $C_{1,2}(\mu )$ are the Wilson coefficients (WCs) . The momentum of $B$ meson, $D^{(*)}$ meson and the lightest meson $\pi$ are denoted as $P_1$, $P_2$ and $P_3$, respectively. With the light-cone coordinate, the momenta of various mesons are assigned as
\begin{equation}
P_1=\frac{m_B}{\sqrt{2}}(1,1,0_\bot), \quad
P_2=\frac{m_B}{\sqrt{2}}(r^2,1,0_\bot), \quad
P_3=\frac{m_B}{\sqrt{2}}(1-r^2,0,0_\bot),
\end{equation}
At the rest frame of $B$ meson, the light meson moves so fast that $P_3^-$ can be treated as zero. The light valence quarks inside the corresponding
mesons are assigned as:
\begin{equation}
k_1=(x_1P_1^+,0,\mathbf{k}_{1T}) \quad
k_2=(0,x_2P_2^-,\mathbf{k}_{2T}) \quad
k_3=(x_3P_3^+,0,\mathbf{k}_{3T})
\end{equation}
with $x_1$,$x_2$ and $x_3$ as the momentum fraction, and $\mathbf{k}_{iT}$ is the transverse momentum of the quark.

\begin{figure}[phtb]
\begin{center}
\includegraphics[width=1.0\linewidth]{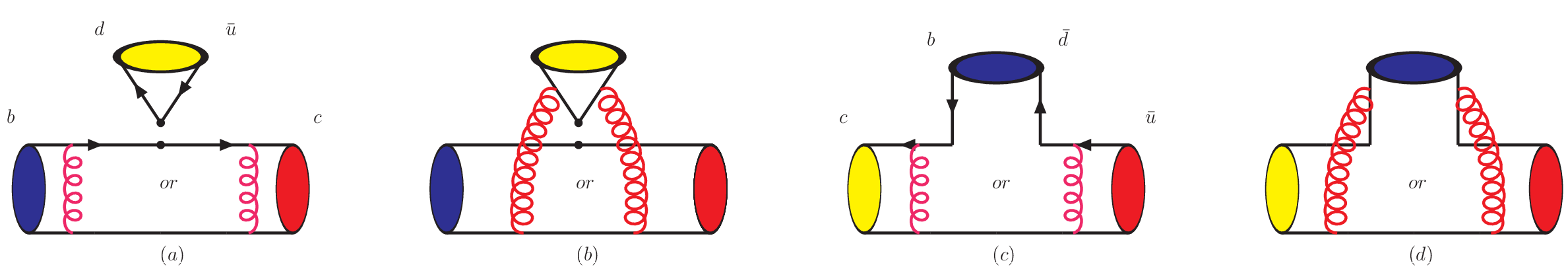}
\caption{ The topologies (a)[(c)] factorizable  emission [annihilation] and (b)[(d)] nonfactorizable effects for the decays $B\to D^{(*)+} \pi^{-}$.} \label{ca_fig} \end{center}
\end{figure}
In terms of the effective interactions, there are two topologies for the decays $\overline B^0 \to D^{(*)+} \pi^-$, emission and annihilation diagrams. The former is color-allowed but the latter belongs to color-suppressed. The corresponding flavor diagrams are illustrated by Fig. \ref{ca_fig}. Hence, the decay amplitude of $\overline B^0 \to D^{(*)+} \pi^-$ can be expressed by
\begin{eqnarray}
{\cal A}(\overline B^0 \to D^{(*)+} \pi^-)= V_{cb}V_{ud}^*\Bigg[M^{D^{(*)}}_{ef} + M^{D^{(*)}}_{enf}+  M^{D^{(*)}}_{af}+ M^{D^{(*)}}_{anf}\Bigg],
\end{eqnarray}
where $M^{D^{(*)}}_{ef}$ and $M^{D^{(*)}}_{af}$ are the contributions of the factorizable emission and annihilation topologies, respectively. $M^{D^{(*)}}_{enf}$ and $M^{D^{(*)}}_{anf}$ denote nonfactorizable contributions. With factorization theorem and  hadronic structures of Eqs. (\ref{waveB}), (\ref{waveD}) and (\ref{wavePI}), the hard amplitudes are formulated as
\begin{align}
M^{D^{(*)}}_{ef} &= 8\pi C_{F} m^{4}_{B} f_{\pi}
\int_{0}^{1} dx_{1} dx_{2}
\int_{0}^{1/\Lambda} b_{1} db_{1} b_{2} db_{2}
\, \phi_B(x_{1},b_{1})\, \phi_{D^{(*)}}(x_{2})
\nonumber\\
& \times \Big\{
(1+r+x_2)\,h_{ef}(x_{1},x_{2},b_{1},b_{2}) \,
{\cal E}_{ef}^{(1)}(t^{(1)}_{ef})  +r\,h_{ef}(x_{2},x_{1},b_{2},b_{1}) \,
{\cal E}_{ef}^{(2)}(t^{(2)}_{ef})
\Big\},
\label{Mef}\\
M^{D^{(*)}}_{enf} &= 16\sqrt{\frac{2}{3}}\pi C_{F} m^{4}_{B}
\int_{0}^{1} dx_{1} dx_{2} dx_{3}
\int_{0}^{1/\Lambda} b_{1} db_{1} b_{3} db_{3}
\, \phi_B(x_{1},b_{1})\, \phi_{D^{(*)}}(x_{2})\, \phi_{\pi}^{A}(x_{3})
\nonumber\\
& \times\Big\{
x_{3}\,h_{enf}^{(1)}(x_{1},x_{2},x_{3},b_{1},b_{3})\,
{{\cal E}_{enf}^{(1)}(t_{enf}^{(1)})}
- (1+x_{2}-x_{3})\,h_{enf}^{(2)}(x_{1},x_{2},x_{3},b_{1},b_{3})\,
{{\cal E}_{enf}^{(2)}(t_{enf}^{(2)})}
\Big\},
\label{Menf}\\
M^{D^{(*)}}_{af} &= 8\pi C_{F} m^{4}_{B} f_{B}
\int_{0}^{1} dx_{2} dx_{3}
\int_{0}^{1/\Lambda} b_{2} db_{2} b_{3} db_{3}
\, \phi_{D^{(*)}}(x_{2})\, \phi_{\pi}^{A}(x_3)
\nonumber\\
& \times\Big\{
-x_{3}\, h_{af}(x_{2},x_{3},b_{2},b_{3})\,
{{\cal E}_{af}^{(1)}(t_{af}^{(1)})}
+ x_{2} \,h_{af}(x_{3},x_{2},b_{3},b_{2})\,
{{\cal E}_{af}^{(2)}(t_{af}^{(2)})}
\Big\},
\label{Maf}\\
M^{D^{(*)}}_{anf} &= 16\sqrt{\frac{2}{3}}\pi C_{F} m^{4}_{B}
\int_{0}^{1} dx_{1} dx_{2} dx_{3}
\int_{0}^{1/\Lambda} b_{1} db_{1} b_{2} db_{2}
\, \phi_B(x_{1},b_{1})\, \phi_{D^{(*)}}(x_{2})\, \phi_{\pi}^A(x_{3})
\nonumber\\
& \times\Big\{
-x_{2}\, h_{anf}^{(1)}(x_{1},x_{2},x_{3},b_{1},b_{2})\,
{{\cal E}_{anf}^{(1)}(t_{anf}^{(1)})}
+ x_{3}\, h_{anf}^{(2)}(x_{1},x_{2},x_{3},b_{1},b_{2})
{{\cal E}_{anf}^{(2)}(t_{anf}^{(2)})}
\Big\}.
\label{Manf}
\end{align}
The hard functions in the amplitude formulas can be written in the following forms:
\begin{align}
h_{ef}(x_1,x_2,b_1,b_2)&=S_{t}(x_{2})K_{0}\left(\sqrt{x_1x_2}m_Bb_1\right)
\nonumber \\
& \times \left[\theta(b_1-b_2)K_0\left(\sqrt{x_2}m_B
b_1\right)I_0\left(\sqrt{x_2}m_Bb_2\right)\right.
\nonumber \\
& \left.+\theta(b_2-b_1)K_0\left(\sqrt{x_2}m_Bb_2\right)
I_0\left(\sqrt{x_2}m_Bb_1\right)\right]\;, \label{dh}
\\
h^{(j)}_{enf}(x_{1},x_{2},x_{3}, b_{1},b_{3})&=
\left[\theta(b_1-b_3)K_0\left(\sqrt{x_{1}x_{2}} m_B
b_1\right)I_0\left(\sqrt{x_{1}x_{2}} m_Bb_3\right)\right. \nonumber \\
& \quad \left. +\theta(b_3-b_1)K_0\left(\sqrt{x_{1}x_{2}}m_B
b_3\right) I_0\left(\sqrt{x_{1}x_{2}}m_B b_1\right)\right]
 \nonumber \\
&  \times \left( \begin{array}{cc}
 K_{0}(D_{j}m_Bb_{3}) &  \mbox{for $D^2_{j} \geq 0$}  \\
 \frac{i\pi}{2} H_{0}^{(1)}(\sqrt{|D_{j}^2|}m_Bb_{3})  &
 \mbox{for $D^2_{j} \leq 0$}
  \end{array} \right),
\label{hjd}\\
h_{af}(x_2, x_3, b_2, b_3) &= S_t(x_3)\left(\frac{i\pi}{2}\right)^2H_0^{(1)}\left(\sqrt{x_2x_3(1-r^2)}m_Bb_2\right)\nonumber\\
&\quad \times \Big[ \theta(b_2 - b_3)J_0(\sqrt{x_3(1-r^2)}m_Bb_3) H_0^{(1)}\left(\sqrt{x_3(1-r^2)}m_Bb_2\right) \nonumber \\
&\quad + \theta(b_3 - b_2)J_0(\sqrt{x_3(1-r^2)}m_Bb_2)H_0^{(1)}\left(\sqrt{x_3(1-r^2)}m_Bb_2\right) \Big],
\\
h^{(j)}_{anf}(x_{1},x_{2},x_{3}, b_{1},b_{2})&= i\frac{\pi}{2}
\left[\theta(b_1-b_2)H_0^{(1)}\left(\sqrt{x_{2}x_{3}(1-r^2)}m_B
b_1\right)J_0\left(\sqrt{x_{2}x_{3}(1-r^2)}m_Bb_2\right)\right. \nonumber \\
&\quad\left.
+\theta(b_2-b_1)H_0^{(1)}\left(\sqrt{x_{2}x_{3}(1-r^2)}m_B
b_2\right)
J_0\left(\sqrt{x_{2}x_{3}(1-r^2)}m_B b_1\right)\right]\;  \nonumber \\
&\times \left( \begin{array}{cc}
 K_{0}(F_{j}m_Bb_{1}) &  \mbox{for $F^2_{j} \geq 0$}  \\
 \frac{i\pi}{2} H_{0}^{(1)}(\sqrt{|F_{j}^2|}m_Bb_{1})  &
 \mbox{for $F^2_{j} \leq 0$}
  \end{array} \right),
\end{align}
with
\begin{align}
D_1^2&=x_1x_2-x_2x_3(1-r^2), \\
D_2^2&=x_1x_2-x_2(1-x_3)(1-r^2),\\
F_1^2&=1-(1-x_2)(1-x_1-x_3(1-r^2)), \\
F_2^2&=x_2(x_1-x_3(1-r^2)).
\end{align}
The evolution factors  are defined by
\begin{align}
{\cal E}^{(i)}_{ef}(t^{(i)}_{ef})&=\left(C_2+\frac{C_1}{3}\right) \alpha_{s}(t^{(i)}_{ef})
\exp\left[-S_{B}(t^{(i)}_{ef})-S_{D^{(*)}}(t^{(i)}_{ef})\right], \\
{\cal E}^{(i)}_{enf}(t^{(i)}_{enf})&=C_1 \alpha_{s}(t^{(i)}_{enf})
\exp\left[-S_{B}(t^{(i)}_{enf})-S_{D^{(*)}}(t^{(i)}_{enf})-S_{\pi}(t^{(i)}_{enf})\right]_{b_2=b_1},
\\
{\cal E}^{(i)}_{af}(t^{(i)}_{af})&=\left(C_1+\frac{C_2}{3}\right) \alpha_{s}(t^{(i)}_{ef})
\exp\left[-S_{D^{(*)}}(t^{(i)}_{af})-S_{\pi}(t^{(i)}_{af})\right], \\
{\cal E}^{(i)}_{anf}(t^{(i)}_{anf})&=C_2 \alpha_{s}(t^{(i)}_{anf})
\exp\left[-S_{B}(t^{(i)}_{anf})-S_{D^{(*)}}(t^{(i)}_{anf})-S_{\pi}(t^{(i)}_{anf})\right]_{b_{3}=b_{2}},
\end{align}
where the exponents $S_{M}(M=B, D^{(*)}, \pi$) are the Sudakov factors. From the above equations, it is evident that the emission contributions are color-allowed and determined by the effective coupling $C_{2}+C_{1}/3$, whereas the annihilation contributions are color-suppressed and governed by $C_{1}+C_{2}/3$. The quantities $t^{(i)}_{ef,enf,af,anf}$ represent the hard scales associated with the corresponding diagrams, which are expected to be of order ${\cal O}(\sqrt{\bar{\Lambda} m^{2}_{B}})\sim 1.6$ GeV on average. The criteria for determining these scales are adopted as
\begin{align}
t_{ef}^{(1)}&=\max(\sqrt{x_2}m_B,1/b_1,1/b_2), \nonumber\\
t_{ef}^{(2)}&=\max(\sqrt{x_1}m_B,1/b_1,1/b_2), \nonumber\\
t_{enf}^{(j)}&=\max(\sqrt{x_1x_2}m_B,\sqrt{ \left | D_j^2 \right | }m_B,1/b_1,1/b_3), \nonumber\\
t_{af}^{(1)}&=\max(\sqrt{x_3(1-r^2)}m_B,1/b_2,1/b_3), \nonumber\\
t_{af}^{(2)}&=\max(\sqrt{x_2(1-r^2)}m_B,1/b_2,1/b_3), \nonumber\\
t_{anf}^{(j)}&=\max(\sqrt{x_2x_3(1-r^2)}m_B,\sqrt{ \left | F_j^2 \right | } m_B,1/b_1,1/b_2). \label{scale}
\end{align}
Since we treat the hadronic effects in $B$ decays by considering six-quark interactions simultaneously, at the lowest order in the strong interaction, in addition to the renormalization group   running from the $m_{W}$ scale down to the $m_{B}$ scale in the $\mu$-dependence of the WCs, it is also necessary to account for the running from the $m_{B}$ scale to the hard scales $t^{(i)}_{ef,enf,af,anf}$, which effectively determine the dynamics of $B$ meson decays. Consequently, in our framework, the hard scales for the WCs are determined according to Eq.~(\ref{scale}), rather than being fixed at $m_{B}$ or $m_{B}/2$. As shown in Eqs. (\ref{Mef})-(\ref{Manf}), the amplitudes depend solely on the pion's twist-2 distribution amplitude, $\phi_\pi^A(x)$.  We also note that in the amplitudes the terms proportional to $r^{2}$ (for right-handed gluon exchange) and to $r_{c}$ (for left-handed gluon exchange), as shown in Fig.\ref{ca_fig}, are neglected. Since the leading-power contributions are not suppressed by $1/m_{B}$, these terms are identified as higher-power corrections.

\begin{table}[htb]
\caption{ The values of hard amplitudes (in units of $10^{-3}$) with fixing $C_{D^{(*)}}=0.80$. } \label{vb}
\begin{center}
\begin{tabular}{|c|c|c|c|c|c|c|}
\hline
Decay Mode& $\omega_{B}({\rm GeV})$ &$M_{ef}$ & $M_{enf}$& $M_{af}$ & $M_{anf}$ & BF \\
\hline
\multirow{3}{*}{$\overline B^0 \to D^{+} \pi^-$}&
$0.35$
& $-97.93 $
& $2.43 -i6.60  $
& $0.15 +i0.22 $
& $1.21 +i8.46 $
& $4.61$ \\ \cline{2-7}
\multirow{3}{*}{}&
$0.40$
&$-78.09 $
&$1.84 -i5.01  $
& $0.15 +i0.22 $
& $1.19  +i7.84 $
& $2.92$ \\ \cline{2-7}
\multirow{3}{*}{}&
$0.45$
&$-63.27 $
&$1.45 -i3.90 $
& $0.15 +i 0.22 $
& $1.22  +i7.37 $
& $1.91$  \\ \hline
\multirow{3}{*}{$\overline B^0 \to D^{*+} \pi^-$}&
$0.35$
&$-108.46  $
&$2.65  -i 7.24  $
& $0.18  +i0.26  $
& $1.41  +i 9.26  $
& $5.52 $ \\ \cline{2-7}
\multirow{3}{*}{}&
$0.40$
&$-86.43 $
&$2.05 -i5.47$
& $0.18 +i0.26 $
& $1.37 +i8.64 $
& $3.49$ \\ \cline{2-7}
\multirow{3}{*}{}&
$0.45$
&$-69.96 $
&$1.61 -i4.23  $
& $0.18+i 0.26$
& $1.36 +i 8.13$
& $2.28$ \\ \hline
\end{tabular}
\end{center}
\end{table}

\begin{table}[htb]
\caption{ The values of hard amplitudes (in units of $10^{-3}$)
with fixing $\omega_{B}=0.40 {\rm GeV}$. } \label{vd}
\begin{center}
\begin{tabular}{|c|c|c|c|c|c|c|}
\hline
Decay Mode& $C_{D^{(*)}}$ &$M_{ef}$ & $M_{enf}$& $M_{af}$ & $M_{anf}$ & BF \\
\hline
\multirow{3}{*}{$\overline B^0 \to D^{+} \pi^-$}
&$0.75$
& $-76.67$
& $1.82-i4.93 $
& $0.14+i0.22$
& $1.13+i7.68$
& $2.82$ \\ \cline{2-7}
\multirow{3}{*}{}&
$0.80$
&$-78.09$
&$1.84-i5.01 $
& $0.15+i0.22$
& $1.19 +i7.84$
& $2.92$ \\
 \cline{2-7}
 \multirow{3}{*}{}&
$0.85$
&$-79.52$
&$1.87-i5.09  $
& $0.16+i0.22$
& $1.26 +i8.01$
& $3.03$\\ \hline
\multirow{3}{*}{$\overline B^0 \to D^{*+} \pi^-$}&
$0.75$
&$-84.86$
&$2.03-i5.38  $
& $0.17+i0.26$
& $1.30 +i 8.46$
& $3.37$  \\ \cline{2-7}
\multirow{3}{*}{}&
$0.80$
&$-86.43$
&$2.05-i5.47  $
& $0.18+i0.26$
& $1.37 +i 8.64$
& $3.49$ \\ \cline{2-7}
\multirow{3}{*}{}&
$0.85$
&$-88.00$
&$2.07-i5.56 $
& $0.19+i 0.25$
& $1.44 +i8.83$
& $3.62$ \\ \hline
\end{tabular}
\end{center}
\end{table}

In our calculation, we adopt $V_{cb}=0.04161$ and  $V_{ud}=0.97385$ for the CKM matrix elements \cite{ParticleDataGroup:2024cfk}. Based on the derived formulas and the adopted meson distribution amplitudes, the magnitudes of the hard amplitudes are presented in Tables.~\ref{vb} and \ref{vd}. Theoretical uncertainties primarily originate from the variation of the shape parameters in the meson distribution amplitudes: $0.35 {\rm GeV} < \omega_B < 0.45 {\rm GeV}$ for the $B$-meson and $0.75 < C_{D^{(*)}} < 0.85$ for the $D^{(*)}$-meson, respectively. The results indicate that the factorizable amplitude $M_{ef}$ provides the dominant contribution. By contrast, the nonfactorizable amplitude $M_{enf}$ is negligible: because the pion distribution amplitude is symmetric under the exchange $x_{3}\leftrightarrow 1-x_{3}$, the contributions from the two diagrams in Figs.~\ref{ca_fig}(c) and \ref{ca_fig}(d) cancel in the dominant region of small $x_{2}$. In addition, it is further suppressed by the small WC $C_{1}$. Furthermore, both factorizable and nonfactorizable annihilation contributions are found to be too small to significantly affect the total amplitudes. Finally, we obtain the theoretical predictions of the branching fractions of $\overline{B}^0 \to D^{(*)+}\pi$ as follow,
\begin{eqnarray}
&{\cal B}(\overline B^0 \to D^{+} \pi^-)  = (2.92^{+1.69+0.10}_{-1.01-0.11})\times 10^{-3}, \\
&{\cal B}(\overline B^0 \to D^{*+} \pi^-) = (3.49 ^{+2.03+0.13 }_{-1.21 -0.12})\times 10^{-3},
\end{eqnarray}
which are consistent with previous calculations \cite{Keum:2003js,Li:2008ts,Zou:2009zza}, with minor deviations attributable to differing choices of nonperturbative parameters like $\omega_b$ and $C_{D^{(*)}}$. 

In comparing with the precise experimental measurements \cite{ParticleDataGroup:2024cfk},
\begin{eqnarray}
&{\cal B}(\overline B^0 \to D^{+} \pi^-) = (2.51\pm 0.08)\times 10^{-3}, \\
&{\cal B}(\overline B^0 \to D^{*+} \pi^-) = (2.66\pm 0.07)\times 10^{-3},
\end{eqnarray}
we find that our theoretical predictions, while compatible within uncertainties, display a systematic tendency for larger central values. Given the inverse correlation between the predicted branching fractions and the shape parameter $\omega_B$, these hadronic decay data favor a larger value of $\omega_B$ than the default $0.4~{\rm GeV}$. This trend presents a notable contrast with the semileptonic $\overline{B}^0 \to D^{(*)+}\ell^-\bar{\nu}_{\ell}$ decays, where data favor a smaller $\omega_B$. The channel dependence of this preferred value underscores the challenge of achieving a universal parameterization of non-perturbative QCD effects. The superior precision of the experimental results sets a clear benchmark for future theoretical work, necessitating reduced uncertainties to enable sharper tests of the SM and to enhance sensitivity to potential NP.

It should be noted that in calculating the $\overline{B}^0 \to D^+\pi$ decay amplitude within the PQCD approach at leading power, several theoretical uncertainties must be taken into account. A primary source arises from the nonperturbative LCDAs of the $B$, $D^{(*)}$, and $\pi$ mesons. Their shapes and parameters, such as the $B$-meson shape parameter $\omega_B$ and the Gegenbauer moments of the pion LCDA, are not precisely known, and variations in these inputs affect the predicted decay rates. Additional uncertainty comes from the charm-quark mass and HQET parameters relevant to the $D$-meson LCDA. The calculation is performed at leading order in $\alpha_s$ and leading power in $\Lambda_{\text{QCD}}/m_b$, so neglected higher-order and power-suppressed effects introduce systematic limitations. In particular, next-to-leading-order corrections and $1/m_b$ terms could modify both the normalization and the scale dependence of the amplitude. The choice of hard scales in PQCD also carries ambiguity, and variations are used to estimate missing higher-order contributions. Moreover, although PQCD assumes that final-state interactions are suppressed by Sudakov effects, residual long-distance rescattering between the $D$ and $\pi$ mesons may still contribute, leading to uncertainties not fully captured in the framework. Sudakov suppressions and threshold resummations themselves depend on specific parameterizations, which further influence numerical outcomes. Finally, uncertainties from CKM inputs, especially $V_{cb}$, and possible violations of the factorization assumption also affect the predictions. Looking ahead, reducing these uncertainties will require more precise nonperturbative inputs, systematic higher-order PQCD calculations, explicit assessments of FSIs, and complementary results from lattice QCD.

\section{Correlation Between $\overline{B}^0 \to D^{(*)+}\ell^-\bar{\nu}_\ell$ and $ \overline{B}^0 \to D^{(*)+}\pi^-$ Decays} \label{sec:correlation}
We analyze both semileptonic decays $\overline{B}^0 \to D^{(*)+}\ell^-\bar{\nu}_\ell$ and nonleptonic decays $\overline{B}^0 \to D^{(*)+}\pi^-$ within the PQCD framework. While PQCD has proven successful in describing many heavy-flavor processes, predictions for the absolute branching fractions of these channels are subject to sizable theoretical uncertainties. These uncertainties originate primarily from nonperturbative inputs, including the  LCDAs of the $B$, $D^{(*)}$, and $\pi$ mesons. Further contributions arise from the choice of hard scales, truncation of the perturbative expansion at leading order in $\alpha_s$, neglected power corrections of order $1/m_b$, and potential final-state interactions in nonleptonic decays. Taken together, these effects limit the predictive power of absolute branching fractions, making them insufficiently precise for stringent tests of the SM or for identifying possible contributions from NP.

To address these challenges, we introduce the ratio
\begin{equation} \label{Rpiell}
R^{(*)}_{\pi/\ell}(q^2)\equiv \frac {\Gamma(\overline{B}^0 \to D^{(*)+} \pi^-)} {d\Gamma(\overline{B}^0 \to D^{(*)+} \ell^- \bar{\nu}_{\ell})/dq^2},
\end{equation}
which compares the nonleptonic branching fraction with the differential semileptonic decay rate. This observable offers several advantages. Because both numerator and denominator involve the same $B\to D^{(*)}$ transition, the dominant hadronic uncertainties largely cancel. Variations in the $B$-meson shape parameter, the LCDA of $D^{(*)}$ meson, and other nonperturbative inputs affect both channels in a similar way, thereby reducing their impact on the ratio. Uncertainties from hard-scale choices, truncation of the perturbative expansion, and resummation parameterizations are also suppressed. Even final-state interactions, which complicate the interpretation of nonleptonic decays, are mitigated since the semileptonic process provides a clean reference. Moreover, the dependence on CKM parameters, particularly $V_{cb}$, cancels almost completely, making $R^{(*)}_{\pi/\ell}(q^2)$ a theoretically robust observable.

\begin{figure}[phtb]
    \centering
         \includegraphics[width=0.48\linewidth]{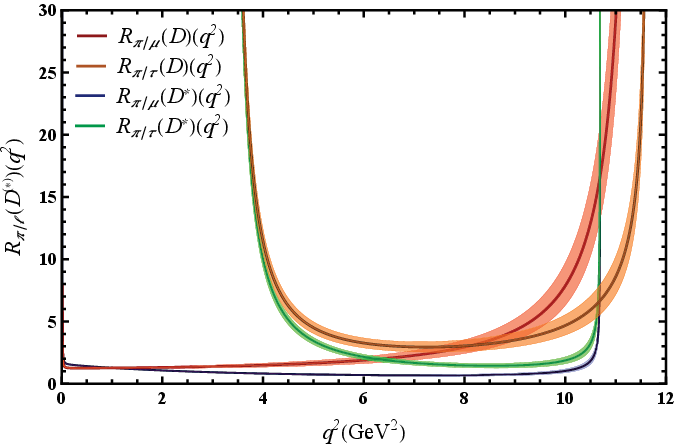}
    \caption{ PQCD predictions for the ratios $R^{(*)}_{\pi/\ell}(q^2)$ defined in Eq.~\eqref{Rpiell}, and the shaded bands in the figure correspond to residual theoretical uncertainties, dominated by higher-order QCD corrections, input LCDAs, and subleading power effects. }
    \label{fig:side:R}
\end{figure}

The differential ratios $R_{\pi/\ell}(q^2)$ and $R^*_{\pi/\ell}(q^2)$, illustrated in Fig.~\ref{fig:side:R}, provide a transparent probe of the interplay between semileptonic form factors and nonleptonic amplitudes. Their shapes reflect the underlying $q^2$ dependence of the semileptonic transitions, while the normalization absorbs much of the hadronic uncertainty. The shaded bands in the figure correspond to residual theoretical uncertainties, dominated by higher-order QCD corrections, input LCDAs, and subleading power effects. Importantly, these uncertainties are far smaller than those associated with absolute branching ratios, thereby significantly enhancing the predictive capacity of the PQCD approach.

From a theoretical perspective, the stability of $R^{(*)}_{\pi/\ell}(q^2)$ across a broad range of $q^2$ values suggests that it serves as a sensitive probe of the factorization hypothesis in nonleptonic $B$ decays. Any notable deviation between predicted and measured distributions-whether in shape or normalization-would be highly informative. Such discrepancies could signal a breakdown of factorization, reveal nonfactorizable QCD dynamics, or point to contributions from NP operators, such as non-standard charged currents. The differential dependence on $q^2$ provides additional diagnostic power, since possible NP effects may alter the distribution in a manner distinct from the SM expectation.

From an experimental standpoint, both semileptonic and nonleptonic modes are accessible with high precision at Belle II and LHCb. Large-statistics measurements of $\overline{B}^0 \to D^{(*)+}\ell^-\bar{\nu}_\ell$ are already available, and improved determinations of $\overline{B}^0 \to D^{(*)+}\pi^-$ continue to be refined with growing datasets. By combining these results, experimentalists can directly construct $R^{(*)}_{\pi/\ell}(q^2)$ either differentially or in integrated form. The cancellation of dominant theoretical uncertainties ensures that any discrepancy between data and PQCD predictions would provide a clean signal of nonfactorizable effects or NP.

In summary, $R^{(*)}_{\pi/\ell}(q^2)$ constitutes a theoretically clean and experimentally accessible observable that reduces hadronic and parametric uncertainties while retaining strong sensitivity to both the SM and potential NP. Its differential $q^2$ dependence allows for detailed tests of factorization and robust comparisons with future high-precision data. With Belle II and LHCb expected to reach percent-level precision in the coming years, this ratio will provide stringent benchmarks for PQCD predictions and deepen our understanding of heavy-quark dynamics in $B$ decays.

\section{Summary} \label{sec:summary}
We have carried out a comprehensive analysis of the semileptonic decays $\overline{B}^0 \to D^{(*)+}\ell^-\bar{\nu}_\ell$ and the nonleptonic Decays $\overline{B}^0 \to D^{(*)+}\pi^-$ within the PQCD framework. For the semileptonic transitions, the form factors are calculated in the low-$q^2$ region, where the large-recoil dynamics are dominated by hard gluon exchange. The resulting PQCD form factors are extrapolated to the high-$q^2$ region using lattice QCD inputs, yielding a consistent description over the full kinematic range. This combined treatment provides an improved theoretical basis for the SM predictions of the LFU observables $R(D)$ and $R(D^*)$, which are found to be compatible with the latest experimental data. In addition, we have examined the differential ratios $R_{D^{(*)}}(q^2)$, which provide a more detailed probe of LFU and will serve as valuable observables in future precision tests. 

For the nonleptonic $\overline{B}^0 \to D^{(*)+}\pi^-$ decays, both factorizable and nonfactorizable contributions are calculated within the same PQCD framework. The shared heavy-to-heavy transition currents between semileptonic and nonleptonic channels allow us to test the validity of factorization and to explore correlations between the two classes of decays. Our results indicate that a unified PQCD treatment provides a coherent and quantitatively reliable description of heavy-to-heavy transitions, reducing hadronic uncertainties and searching for effects of new physics beyond the SM.

Future improvements may include incorporating next-to-leading-order PQCD corrections and updated lattice QCD inputs with higher precision at large-$q^2$. Such developments, together with forthcoming high-luminosity data from Belle II and the upgraded LHCb, will enable more stringent tests of LFU and offer enhanced sensitivity to possible new physics in heavy-flavor transitions.

\section*{Acknowledgments}
This work is supported in part by the National Science Foundation of China under the Grants No. 12375089 and 12435004, and the Natural Science Foundation of Shandong province under the Grant No. ZR2022ZD26 and ZR2022MA035.
\bibliographystyle{bibstyle}
\bibliography{Ref}
\end{document}